\documentclass[prd,preprint,preprintnumbers,nofootinbib,eqsecnum,superscriptaddress]{revtex4}

 \usepackage[dvips,final]{graphicx}
  \usepackage{amssymb}
   \usepackage{amsmath}
    \usepackage{amsfonts}
     \usepackage{epsfig}
      \usepackage{bm}

\usepackage{mathpazo}

\usepackage[section]{placeins}

\usepackage{multirow}
\usepackage{ctable}
\usepackage{booktabs}
\usepackage{array}
\usepackage{tabularx}
\usepackage{xcolor}
\usepackage{pstricks}

\begin{document}

\title{Double-parton scattering effects in \bm{$D^{0}B^{+}$} and \bm{$B^{+}B^{+}$} meson-meson pair production in proton-proton collisions at the LHC
}

\author{Rafa{\l} Maciu{\l}a}
\email{rafal.maciula@ifj.edu.pl} \affiliation{Institute of Nuclear
Physics, Polish Academy of Sciences, Radzikowskiego 152, PL-31-342 Krak{\'o}w, Poland}

\author{Antoni Szczurek\footnote{also at University of Rzesz\'ow, PL-35-959 Rzesz\'ow, Poland}}
\email{antoni.szczurek@ifj.edu.pl} \affiliation{Institute of Nuclear
Physics, Polish Academy of Sciences, Radzikowskiego 152, PL-31-342 Krak{\'o}w, Poland}


\begin{abstract}
We extend our previous studies of double-parton scattering (DPS) to
simultaneous production of $c \bar c$ and $b \bar b$ and production of
two pairs of $b \bar b$. The calculation is performed within factorized
ansatz. Each parton scattering is calculated within $k_T$-factorization 
approach. The hadronization is done with the help of fragmentation
functions. Production of $D$ mesons in our framework was tested in 
our previous works.
Here we present our predictions for $B$ mesons. A good agreement is
achieved with the LHCb data.
We present our results for $c \bar c b \bar b$ and $b \bar b b \bar b$
final states. For completeness we compare results for double- and single-parton
scattering (SPS). As for $c \bar c c \bar c$ final state also here
the DPS dominates over the SPS, especially for small transverse momenta.
We present several distributions and integrated cross sections 
with realistic cuts for simultaneous production of $D^0 B^+$ and $B^+
B^+$, suggesting future experimental studies at the LHC.
\end{abstract}


\maketitle

\section{Introduction}

Phenomena of multiple-parton interaction (MPI) have become very important for precise description of high-energy proton-proton collisions in the ongoing LHC era.
There are several experimental and theoretical studies of soft and hard MPI effects in progress (see \textit{e.g.} Refs.~\cite{Astalos:2015ivw,Proceedings:2016tff}), so far mostly concentrated on double-parton scattering (DPS). In many cases exploration of DPS mechanisms for different processes needs dedicated experimental analysis and is strongly limited because of large background coming from standard single-parton scattering (SPS).

Some time ago we proposed and discussed double open charm meson production $pp \to D D \!\; X$ as a potentially one of the best reaction to study hard double-parton scattering effects at the LHC \cite{Luszczak:2011zp}. This conclusion was further confirmed by the LHCb collaboration that has reported surprisingly large cross sections for $DD$ meson-meson pair production in $pp$-scattering at $7$ TeV \cite{Aaij:2012dz}. 
As we have shown in our subsequent studies the LHCb double charm data cannot be explained without the DPS mechanism \cite{Maciula:2013kd}. In this case the standard SPS contribution is much smaller and the data sample is clearly dominated by the DPS component \cite{vanHameren:2014ava,vanHameren:2015wva}.

Subsequently, we have done similar phenomenological studies for other final states. We identified optimal conditions for exploring
DPS effects in $pp \to \mathrm{4jets} \!\; X$ \cite{Maciula:2015vza,Kutak:2016ukc} as well as in $pp \to D^{0} + \mathrm{2jets} \!\; X$ and $pp \to D^{0}\overline{D^{0}} + \mathrm{2jets} \!\; X$ \cite{Maciula:2017egq} reactions for the ATLAS experiment. Very recently, we have also discussed for the first time possible observation of triple-parton scattering (TPS) mechanism in triple open charm meson production with the LHCb detector \cite{Maciula:2017meb}. 
Some rather general features of double-parton scattering were discussed 
previously both for $b \bar b b \bar b$ \cite{DelFabbro:2002pw} and 
$c \bar c b \bar b$ \cite{Cazaroto:2013fua} final states. Here we extend
the discussion by including also single-parton scattering mechanism
for a first time.

In this paper, we wish to present results of phenomenological studies of
DPS effects in the case of associated open charm and bottom 
$pp \to D^{0} B^{^{+}} \!\; X$
as well as double open bottom $pp \to B^{+} B^{+} \!\; X$ production. 
In particular, we will show theoretical predictions of integrated 
and differential cross sections for different energies that could help 
to conclude whether and how the DPS effects for these two cases can 
be observed experimentally by the LHCb/CMS collaborations. 


\section{A sketch of the theoretical formalism}

\subsection{Single-parton scattering}

\begin{figure}[!h]
\begin{minipage}{0.32\textwidth}
 \centerline{\includegraphics[width=1.0\textwidth]{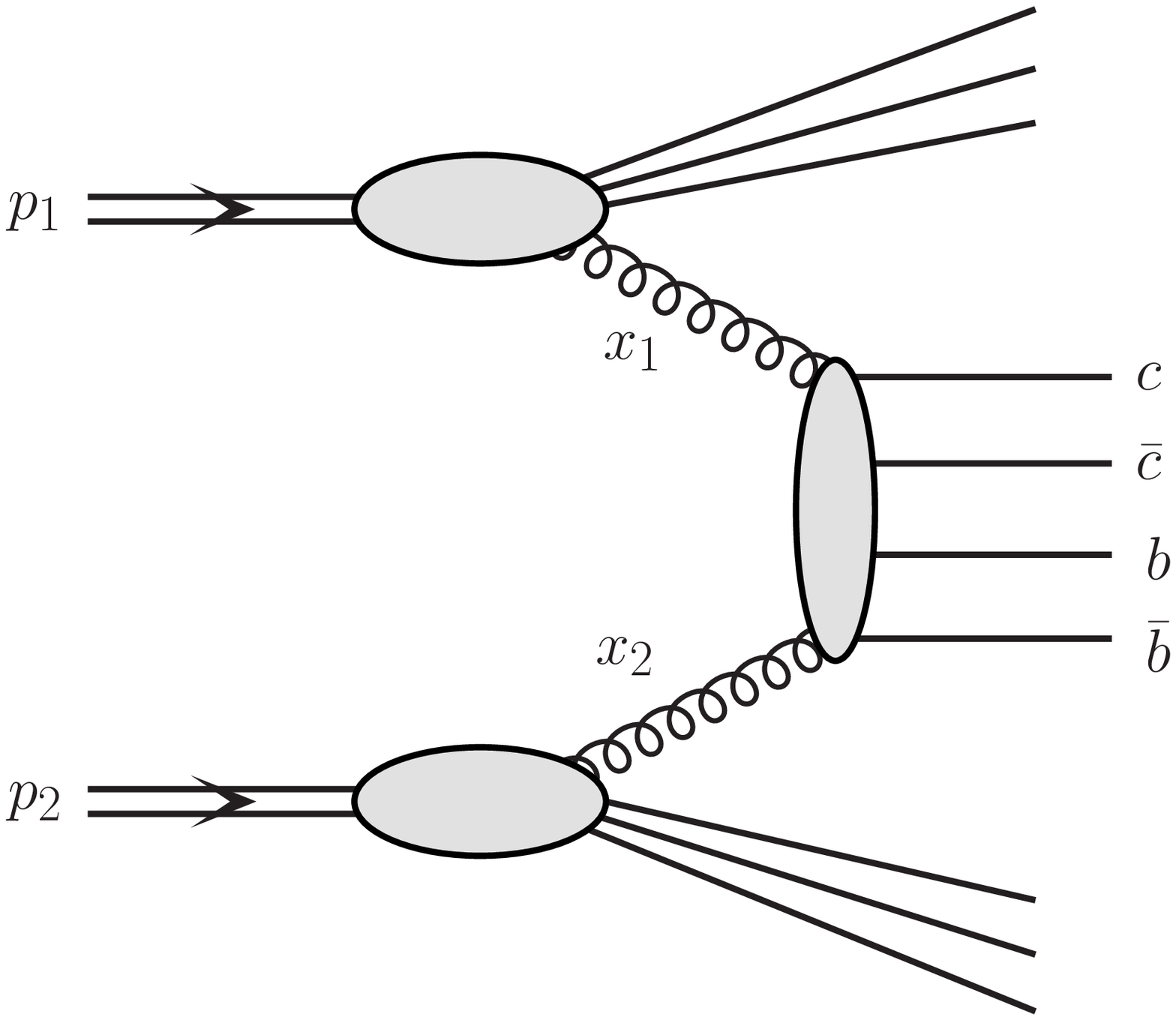}}
\end{minipage}
\hspace{0.1cm}
\begin{minipage}{0.32\textwidth}
 \centerline{\includegraphics[width=1.0\textwidth]{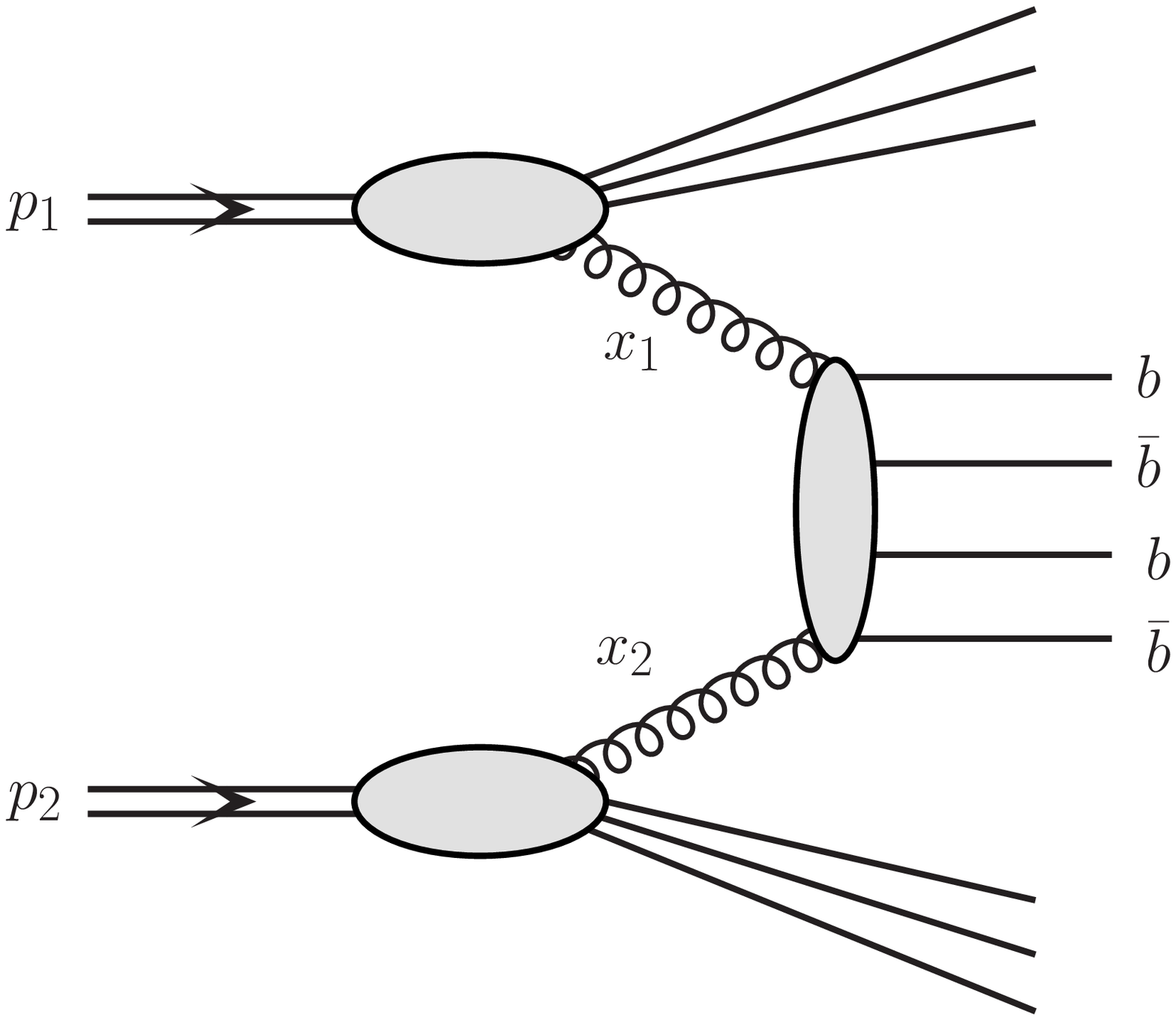}}
\end{minipage}
\caption{A diagrammatic representation of the dominant SPS mechanism
            for the $pp \to c\bar c b\bar b \, X$ (left panel) and for the $pp \to b\bar b b\bar b \, X$ (right panel) reactions.
\small 
 }
 \label{fig:diagrams_SPS}
\end{figure}

In Fig.~\ref{fig:diagrams_SPS} we show a diagrammatic representation of the dominant SPS mechanism for double heavy quark pair production. In particular, in the following we consider mixed $c\bar c b\bar b$ (left panel) and double bottom $b\bar b b\bar b$ (righ panel) final states, however, here the production mechanism is the same as was discussed by us in the case of double charm production (see \textit{e.g.} Ref.~\cite{vanHameren:2015wva}).    

In the $k_T$-factorization approach \cite{Catani:1990xk,Catani:1990eg,Collins:1991ty,Gribov:1984tu} the SPS cross section for $pp \to Q\bar Q Q\bar Q \, X$ reaction can be written as
\begin{equation}
d \sigma_{p p \to Q\bar Q Q\bar Q \; X} =
\int d x_1 \frac{d^2 k_{1t}}{\pi} d x_2 \frac{d^2 k_{2t}}{\pi}
{\cal F}_{g}(x_1,k_{1t}^2,\mu^2) {\cal F}_{g}(x_2,k_{2t}^2,\mu^2)
d {\hat \sigma}_{gg \to Q\bar Q Q\bar Q}
\; .
\label{cs_formula}
\end{equation}
In the formula above ${\cal F}_{g}(x,k_t^2,\mu^2)$ is the unintegrated
gluon distribution function (uGDF). The uGDF depends on longitudinal momentum fraction $x$, transverse momentum squared $k_t^2$ of the gluons entering the hard process,
and in general also on a (factorization) scale of the hard process $\mu^2$.
The elementary cross section in Eq.~(\ref{cs_formula}) can be written
somewhat formally as:
\begin{equation}
d {\hat \sigma}_{gg \to Q \bar Q Q \bar Q } =
\prod_{l=1}^{4}
\frac{d^3 p_l}{(2 \pi)^3 2 E_l} 
(2 \pi)^4 \delta^{4}(\sum_{l=1}^{4} p_l - k_1 - k_2) \times\frac{1}{\mathrm{flux}} \overline{|{\cal M}_{g^* g^* \to Q \bar Q Q \bar Q}(k_{1},k_{2})|^2}
\; ,
\label{elementary_cs}
\end{equation}
where $E_{l}$ and $p_{l}$ are energies and momenta of final state heavy quarks. Above only dependence of the matrix element on four-vectors of incident partons $k_1$ and $k_2$ is made explicit. In general all four-momenta associated with partonic legs enter.
The matrix element takes into account that both gluons entering the hard
process are off-shell with virtualities $k_1^2 = -k_{1t}^2$ and $k_2^2 = -k_{2t}^2$.
In numerical calculations we limit ourselves to the dominant gluon-gluon fusion channel of the $2 \to 4$ type parton-level mechanism.
We checked numerically that the channel induced by the $q\bar q$-annihilation can be safely neglected in the kinematical region under consideration here.

The off-shell matrix elements for higher final state parton
multiplicities, at the tree-level are calculated analytically applying
well defined Feynman rules \cite{vanHameren:2012if} or recursive methods,
like  generalised  BCFW recursion \cite{vanHameren:2014iua}, or
numerically with the help of methods of numerical BCFW recursion 
\cite{Bury:2015dla}. The latter method was already applied for 
$2 \to 3$ production mechanisms in the case of 
$c\bar c + \mathrm{jet}$ \cite{Maciula:2016kkx} and even 
for $2 \to 4$ processes in the case of $c\bar c c\bar c$ 
\cite{vanHameren:2015wva}, four-jet \cite{Kutak:2016mik} and 
$c\bar c + \mathrm{2jets}$ \cite{Maciula:2017egq} final states.

In this paper we use the same numerical methods. The calculation 
is performed with the help of KaTie \cite{vanHameren:2016kkz}, which is 
a complete Monte Carlo parton-level event generator for hadron
scattering processes. 
It can can be applied to any arbitrary processes within the Standard
Model, for several final-state particles, and for any initial partonic
state with on-shell or off-shell partons. The scattering amplitudes 
are calculated numerically as a function of the external four-momenta 
via Dyson-Schwinger recursion \cite{Caravaglios:1995cd} generalized also
to tree-level off-shell amplitudes. The phase space integration is done 
with the help of a Monte Carlo program with an adaptive phase 
space generator, previously incorporated as a part of 
the AVHLIB library ~\cite{vanHameren:2007pt,vanHameren:2010gg}.

In the present calculation, we use $\mu^2 \! = \! \sum_{i=1}^{4} m_{it}^{2}/4$ as the renormalization/factorization scale, where $m_{it}$'s are the transverse masses of the outgoing heavy quarks. We take running $\alpha_{s}$ at next-to-leading order (NLO),
charm quark mass $m_c$ = 1.5 GeV and bottom quark mass $m_b$ = 4.75 GeV. Uncertainties related to the choice of the parameters were
discussed very recently in Ref.~\cite{Maciula:2017egq} and will be not considered here. We use the Kimber-Martin-Ryskin (KMR) \cite{Kimber:2001sc,Watt:2003vf} unintegrated distributions for gluon calculated from the MMHT2014nlo PDFs \cite{Harland-Lang:2014zoa}. The above choices are kept the same also in the case of double-parton scattering calculation except of the scales.  

The effects of the $c \to D^{0}$ and $b \to B^{+}$ hadronization are taken into account via standard fragmentation function (FF) technique.
We use the scale-independent Peterson model of FF \cite{Peterson:1982ak}
with $\varepsilon_{c} = 0.05$ and $\varepsilon_{b} = 0.004$ which is
commonly used in the literature in the context of heavy quark fragmentation.  
Details of the fragmentation procedure together with discussion of the uncertainties related to the choice of the FF model
can be found \textit{e.g.} in Ref.~\cite{Maciula:2013wg}. In the last step, the cross section for meson is normalized by the relevant branching fractions $\mathrm{BR}(c \to D^{0}) = 0.565$ and $\mathrm{BR}(b \to B^{+}) = 0.4$.    

\subsection{Double-parton scattering}

A formal theory of multiple-parton scattering (see \textit{e.g.}
Refs.~\cite{Diehl:2011tt,Diehl:2011yj}) is rather well established but still not fully applicable for phenomenological studies. In general, the DPS cross sections can be expressed in terms of the double parton distribution functions (dPDFs).
However, the currently available models of the dPDFs are still rather 
at a preliminary stage. So far they are formulated only for gluon or for valence quarks and only in a leading-order framework
which is for sure not sufficient for many processes, especially when heavy quark production is considered.    

Instead of the general form, one usually follows the assumption of the factorization of the DPS cross section.
Within the factorized ansatz, the dPDFs are taken in the following form:  
\begin{equation}
D_{1, 2}(x_1,x_2,\mu) = f_1(x_1,\mu)\, f_2(x_2,\mu) \, \theta(1-x_1-x_2) \, ,
\end{equation}
where $D_{1, 2}(x_1,x_2,\mu)$ is the dPDF and
$f_i(x_i,\mu)$ are the standard single PDFs for the two generic partons in the same proton. The factor $\theta(1-x_1-x_2)$ ensures that
the sum of the two parton momenta does not exceed 1. 
%
\begin{figure}[!h]
\begin{minipage}{0.32\textwidth}
 \centerline{\includegraphics[width=1.0\textwidth]{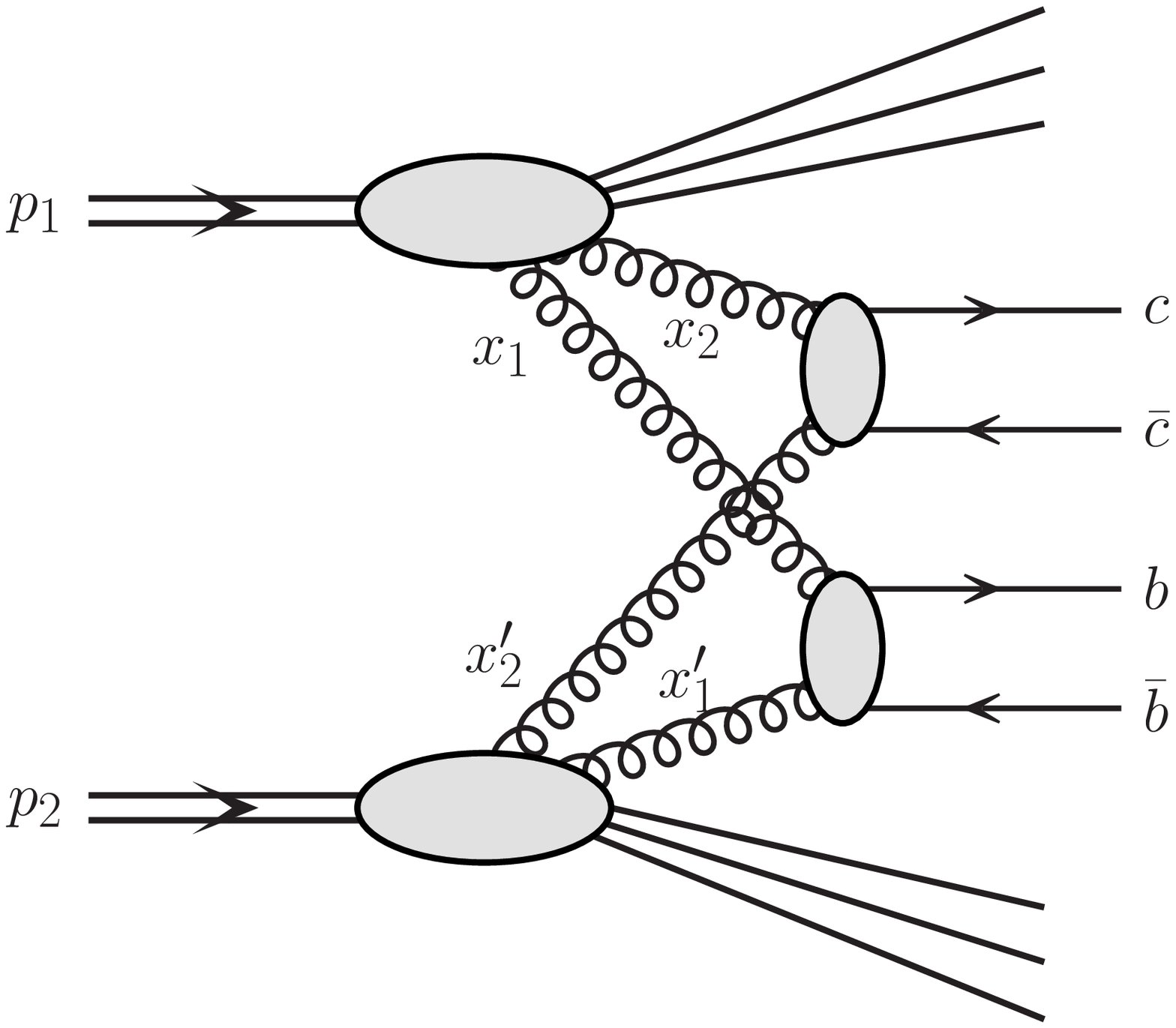}}
\end{minipage}
\hspace{0.1cm}
\begin{minipage}{0.32\textwidth}
 \centerline{\includegraphics[width=1.0\textwidth]{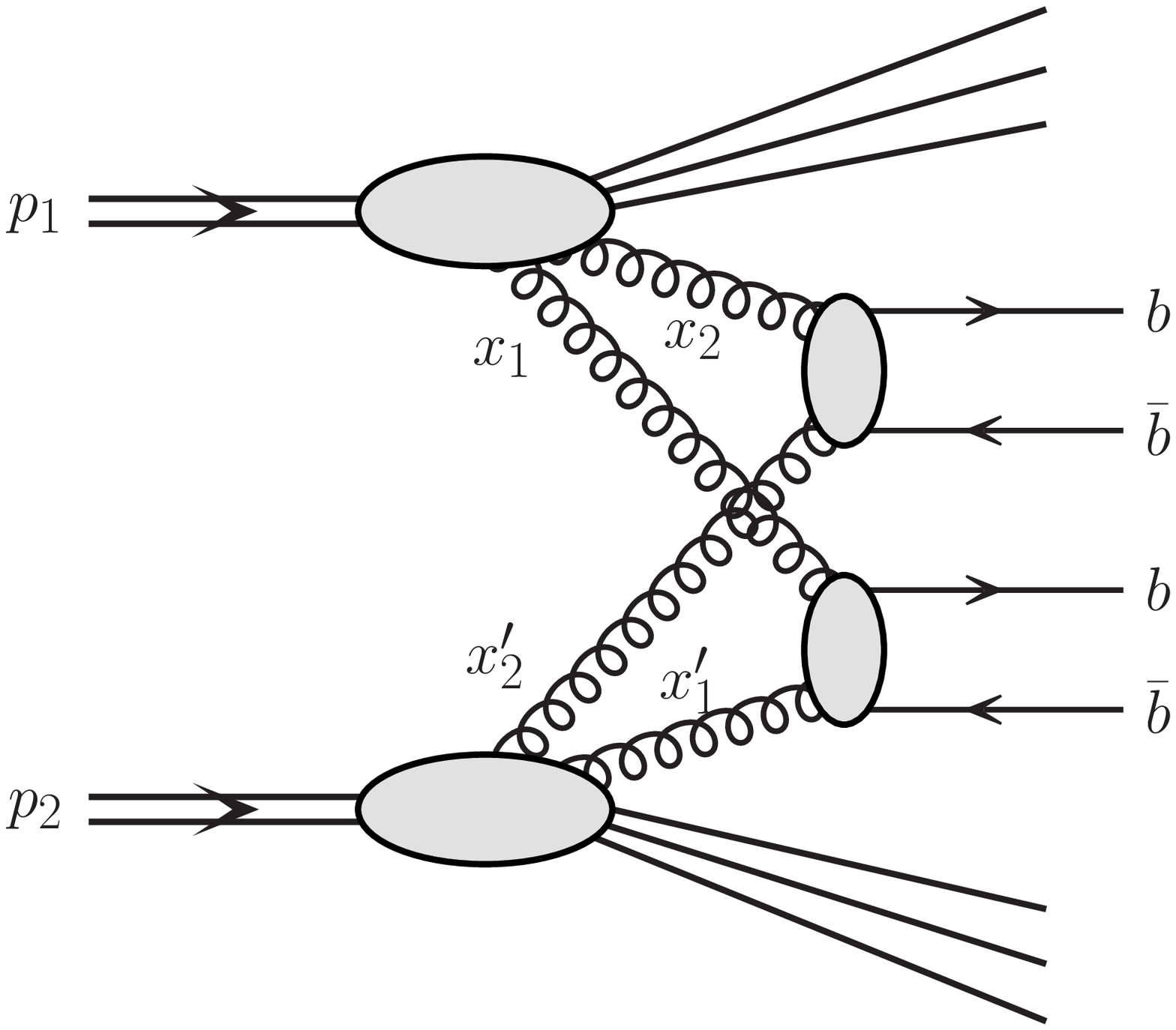}}
\end{minipage}
   \caption{A diagrammatic representation of the DPS mechanism
            for the $pp \to c\bar c b\bar b \, X$ (left panel) and for the $pp \to b\bar b b\bar b \, X$ (right panel) reactions.
\small 
 }
 \label{fig:diagrams_DPS}
\end{figure}

The differential cross section for $pp \to Q \bar Q Q \bar Q \; X$ reaction within the DPS mechanism, sketched in Fig.~\ref{fig:diagrams_DPS}, can be then expressed as follows: 
\begin{equation}
\frac{d\sigma^{DPS}(Q \bar Q Q \bar Q)}{d\xi_{1}d\xi_{2}} =  \frac{m}{\sigma_{\mathrm{eff}}} \cdot \frac{d\sigma^{SPS}(g g \to Q \bar Q)}{d\xi_{1}} \! \cdot \! \frac{\sigma^{SPS}(g g \to Q \bar Q)}{d\xi_{2}},
\label{basic_formula1}
\end{equation}
where $\xi_{1}$ and $\xi_{2}$ stand for generic phase space kinematical variables for the first and second scattering, respectively.
The combinatorial factor $m$ is equal $1$ for $c\bar c b\bar b$ and $0.5$ for $b \bar b b\bar b$ case. 
When integrating over kinematical variables one recovers the commonly used pocket-formula:
\begin{equation}
\sigma^{DPS}(Q \bar Q Q \bar Q) =  
m \! \cdot \!  \frac{\sigma^{SPS}(g g \to Q \bar Q) \! \cdot \! \sigma^{SPS}(g g \to Q \bar Q)}{\sigma_{\mathrm{eff}}}\; .
\label{basic_formula2}
\end{equation}

The effective cross section $\sigma_{\mathrm{eff}}$ provides normalization of 
the DPS cross section and can be roughly interpreted 
as a measure of the transverse correlation of the two partons inside 
the hadrons. The longitudinal parton-parton correlations are far less
important when the energy of the collision is increased, due to the
increase in the parton multiplicity. For small-$x$ partons and for low 
and intermediate scales the possible longitudinal correlations can be safely
neglected (see \textit{e.g.} Ref.~\cite{Gaunt:2009re}). 
In this paper we use world-average value of $\sigma_{\mathrm{eff}} = 15$ mb provided by 
several experiments at Tevatron
\cite{Abe:1997bp,Abe:1997xk,Abazov:2009gc} and LHC
\cite{Aaij:2012dz,Aad:2013bjm,Chatrchyan:2013xxa,Aad:2014rua,Aaboud:2016dea}.
Future experiments may verify this value and establish a systematics.

There are several effects that may lead to a violation of the factorized
ansatz (\ref{basic_formula1}), which seems a priori a severe approximation.
The flavour, spin and color correlations lead, in principle, 
to interference effects that result in breaking the pocket-formula 
(see \textit{e.g.} Refs.~\cite{Diehl:2011tt,Diehl:2011yj}). 
In any case, the spin polarization of the two partons from one hadron
can be mutually correlated, especially when the partons are relatively close in phase space (having comparable $x$'s). The two-parton distributions have a nontrivial color structure which also may lead to a non-negligible correlations effects. 
Such effects are usually not included in phenomenological analyses. They were exceptionally discussed in the context of double charm production \cite{Echevarria:2015ufa} but in this case the corresponding effects were found to be very small.
Moreover, including perturbative parton splitting mechanism \cite{Ryskin:2011kk,Gaunt:2012dd,Gaunt:2014rua} and/or imposing sum rules \cite{Golec-Biernat:2015aza} also leads to a breaking of the pocket-formula.  
However, taken the above and looking forward to further improvements in 
this field, here we limit ourselves to a more pragmatic approach.

In our present analysis cross sections for each step of the DPS
mechanism are calculated in the $k_T$-factorization approach, that is:
\begin{eqnarray}
\frac{d \sigma^{SPS}(p p \to Q \bar Q \; X_1)}{d y_1 d y_2 d^2 p_{1,t} d^2 p_{2,t}} 
&& = \frac{1}{16 \pi^2 {\hat s}^2} \int \frac{d^2 k_{1t}}{\pi} \frac{d^2 k_{2t}}{\pi} \overline{|{\cal M}_{g^{*} g^{*} \rightarrow Q \bar{Q}}|^2} \nonumber \\
&& \times \;\; \delta^2 \left( \vec{k}_{1t} + \vec{k}_{2t} - \vec{p}_{1t} - \vec{p}_{2t}
\right)
{\cal F}_{g}(x_1,k_{1t}^2,\mu^2) {\cal F}_{g}(x_2,k_{2t}^2,\mu^2),
\nonumber
\end{eqnarray}
\begin{eqnarray}
\frac{d \sigma^{SPS}(p p \to Q \bar Q \; X_2)}{d y_3 d y_4 d^2 p_{3,t} d^2 p_{4,t}} 
&& = \frac{1}{16 \pi^2 {\hat s}^2} \int \frac{d^2 k_{3t}}{\pi} \frac{d^2 k_{4t}}{\pi} \overline{|{\cal M}_{g^{*} g^{*} \rightarrow Q \bar Q}|^2} \nonumber \\
&&\times \;\; \delta^2 \left( \vec{k}_{3t} + \vec{k}_{4t} - \vec{p}_{3t} - \vec{p}_{4t}
\right)
{\cal F}_{i}(x_3,k_{3t}^2,\mu^2) {\cal F}_{j}(x_4,k_{4t}^2,\mu^2). \nonumber \\
\end{eqnarray}
The numerical calculations for both SPS mechanisms are also done within the KaTie code, where the relevant fully gauge-invariant off-shell $2 \to 2$ matrix element ${\cal M}_{g^{*} g^{*} \rightarrow Q \bar{Q}}$ is obtained numerically. Its useful analytical form can be found \textit{e.g.} in Ref.~\cite{Catani:1990eg}.  
Here, the strong coupling constant $\alpha_S$ and uGDFs are taken the
same as in the case of the calculation of the SPS mechanism. The
factorization and renormalization scales for the two single scatterings
are $\mu^2 \! = \! \frac{m_{1t}^{2}+m_{2t}^{2}}{2}$ for the first, and 
$\mu^2 \! = \! \frac{m_{3t}^{2}+m_{4t}^{2}}{2}$ for the second subprocess.

\section{Numerical results}

\begin{figure}[!h]
\begin{minipage}{0.47\textwidth}
 \centerline{\includegraphics[width=1.0\textwidth]{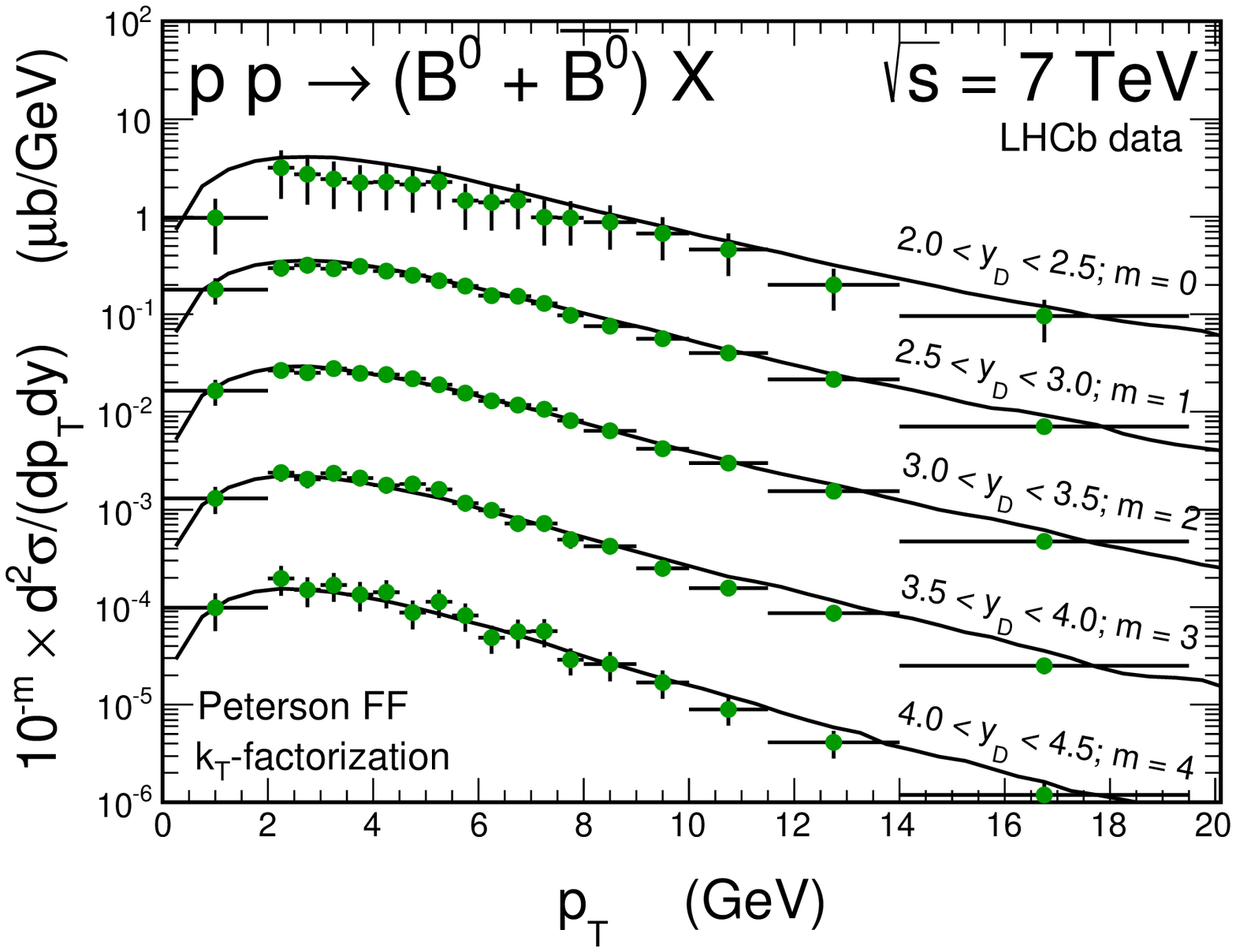}}
\end{minipage}
\hspace{0.5cm}
\begin{minipage}{0.47\textwidth}
 \centerline{\includegraphics[width=1.0\textwidth]{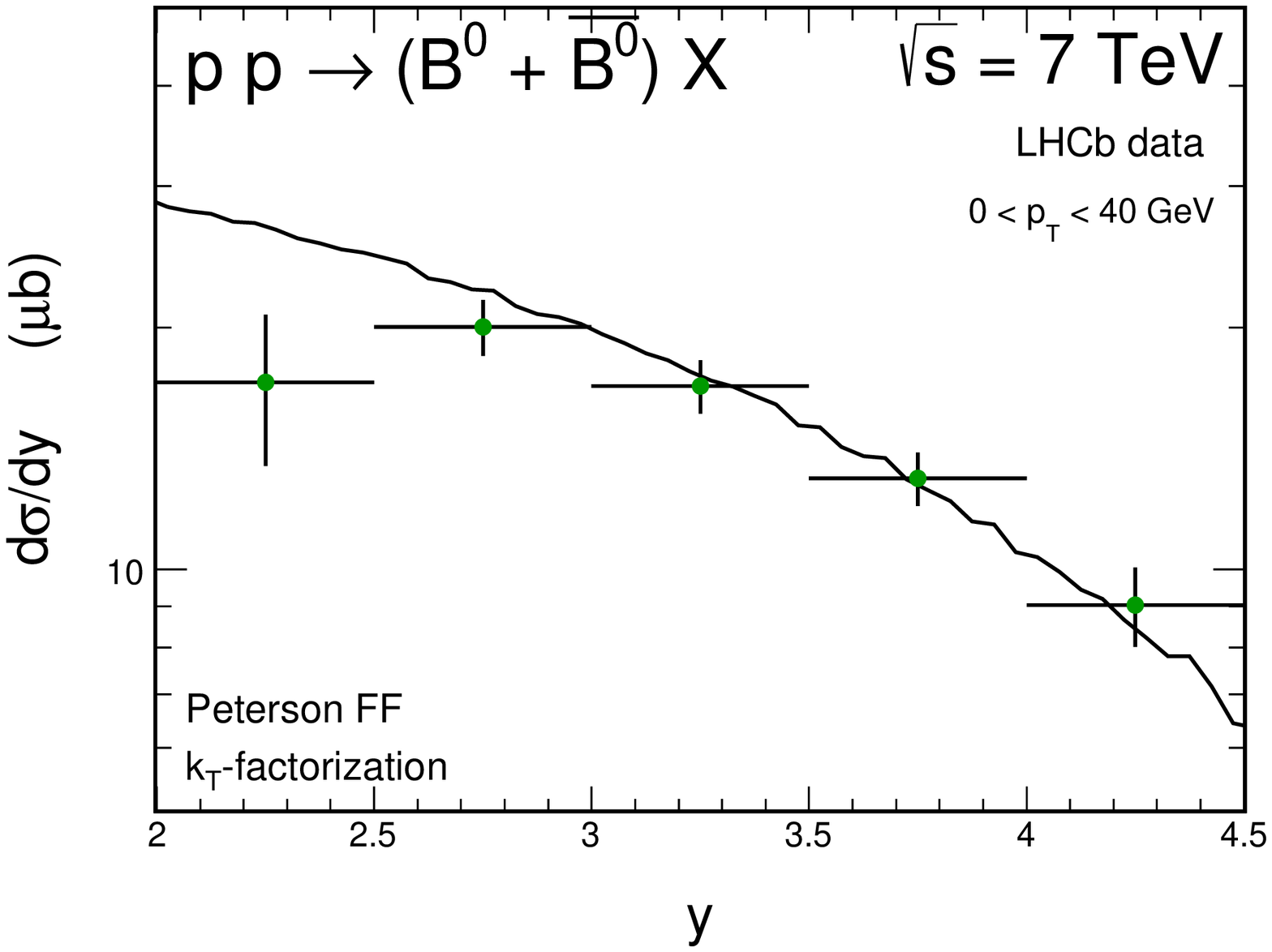}}
\end{minipage}
   \caption{
\small Transverse momentum (left) and rapidity (right) distributions of $B^{0}$ meson measured by the LHCb experiment at $\sqrt{s} = 7$ TeV \cite{Aaij:2013noa}. Theoretical predictions (solid lines) are calculated within the $k_{T}$-factorization approach with the KMR uPDFs. Details are specified in the figure.
 }
 \label{fig:B0}
\end{figure}

Let us start this section with presentation of results of our
calculations for inclusive open bottom meson production. In
Fig.\ref{fig:B0} we compare our theoretical predictions based on the
$k_{T}$-factorization approach with the LHCb experimental data
\cite{Aaij:2013noa} at $\sqrt{s}=7$ TeV. We get a very good agreement
with the experimental points for both, the transverse momentum (left
panel) and rapidity (right panel) $B^{0}$ meson distributions. Only the
cross section in the lowest rapidity bin $y \in (2.0,2.5)$ seems to be
slightly overestimated, however the experimental uncertainties in this
case are noticeably larger than in other rapidity intervals. Similar
high-level agreement between the $k_{T}$-factorization predictions and
experimental data has been also reported by us in the case of inclusive
open charm meson production (see \textit{e.g.}
Ref.~\cite{Maciula:2016wci}). This approach was found to be very
efficient also for more exclusive correlation observables
\cite{Maciula:2013wg,Karpishkov:2016hnx}. Having those conclusions in
mind we expect that the chosen theoretical framework should provide a
reliable predictions also for simultaneous production of charm and
bottom as well as for double bottom production.        

\begin{figure}[!h]
\begin{minipage}{0.47\textwidth}
 \centerline{\includegraphics[width=1.0\textwidth]{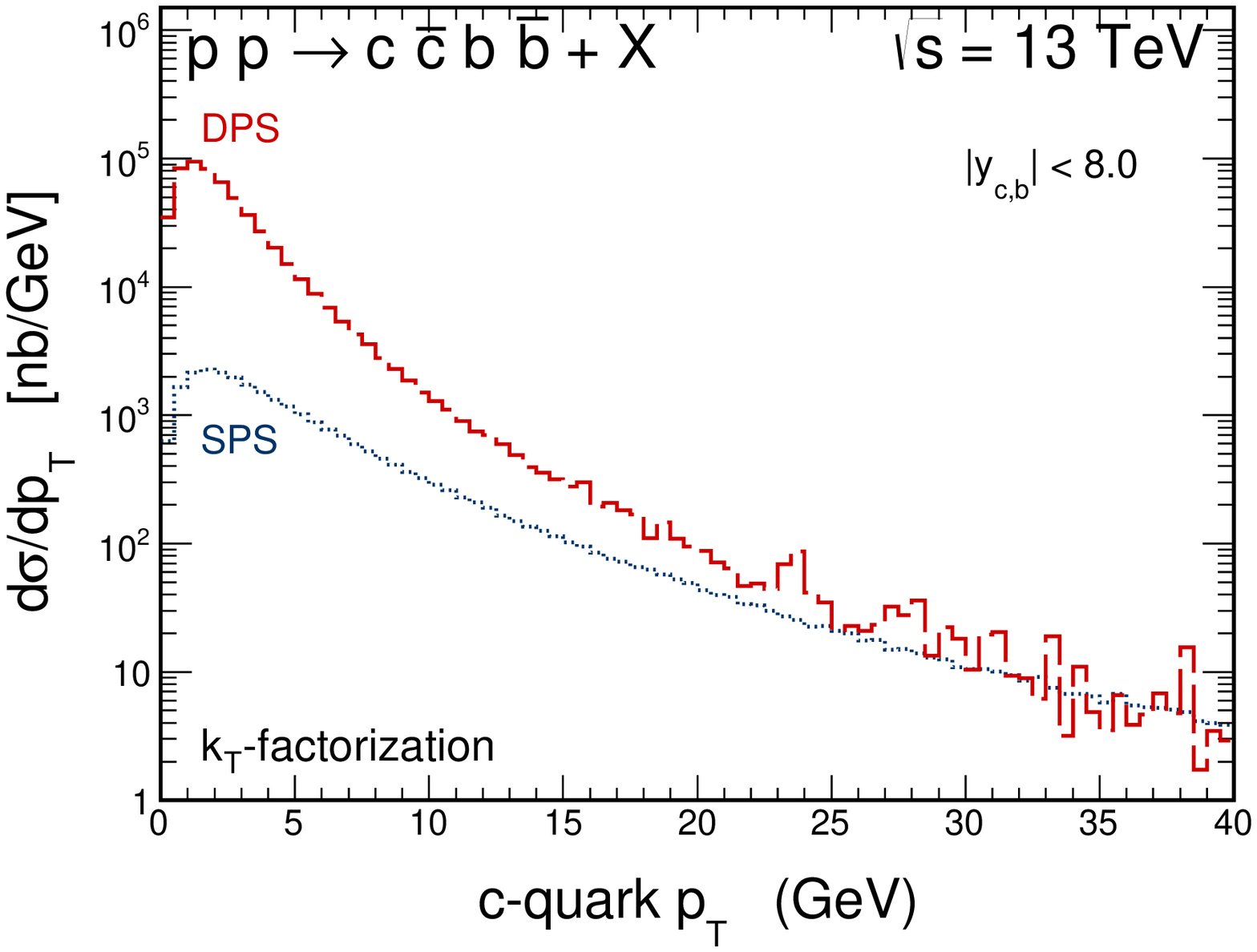}}
\end{minipage}
\hspace{0.5cm}
\begin{minipage}{0.47\textwidth}
 \centerline{\includegraphics[width=1.0\textwidth]{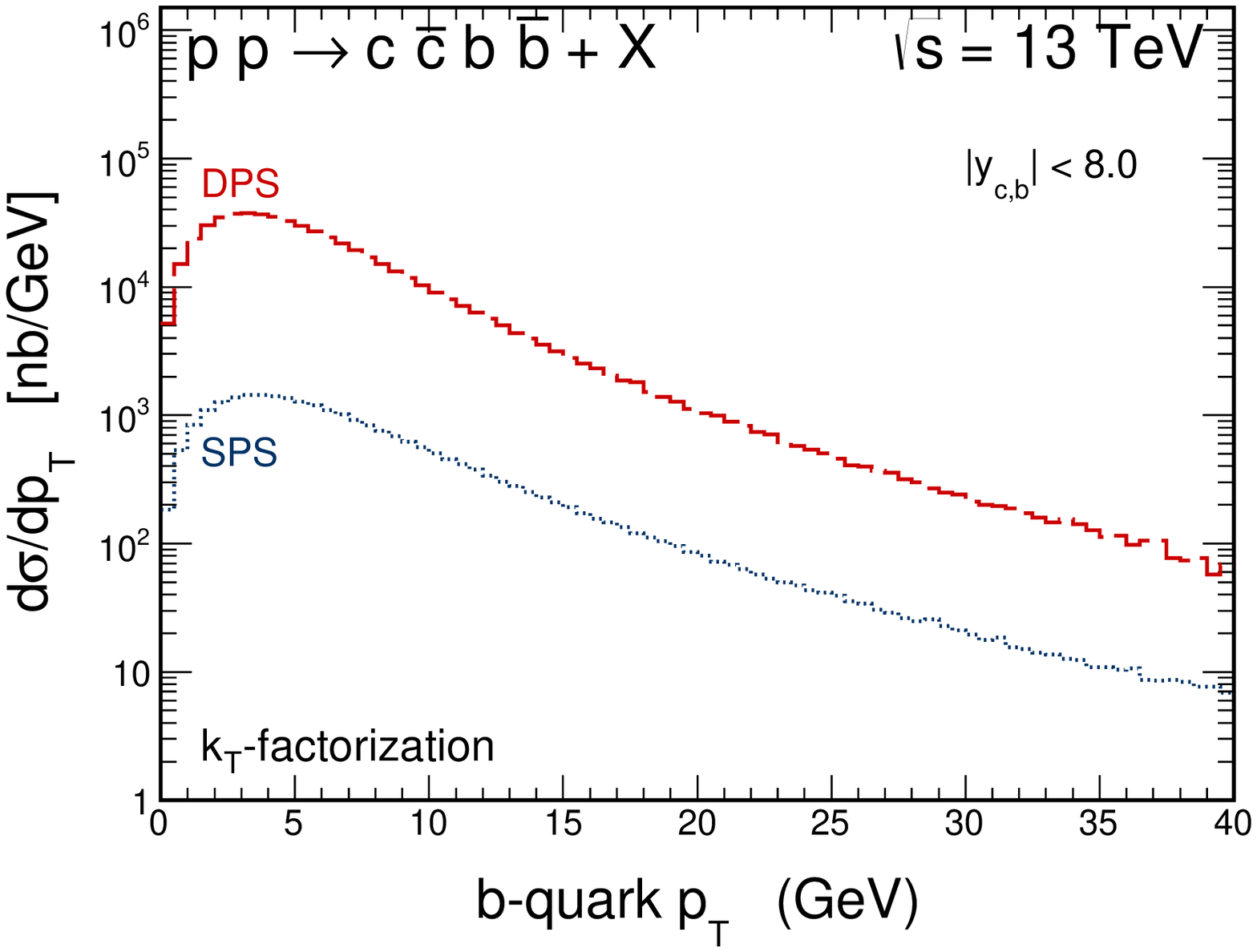}}
\end{minipage}
\begin{minipage}{0.47\textwidth}
 \centerline{\includegraphics[width=1.0\textwidth]{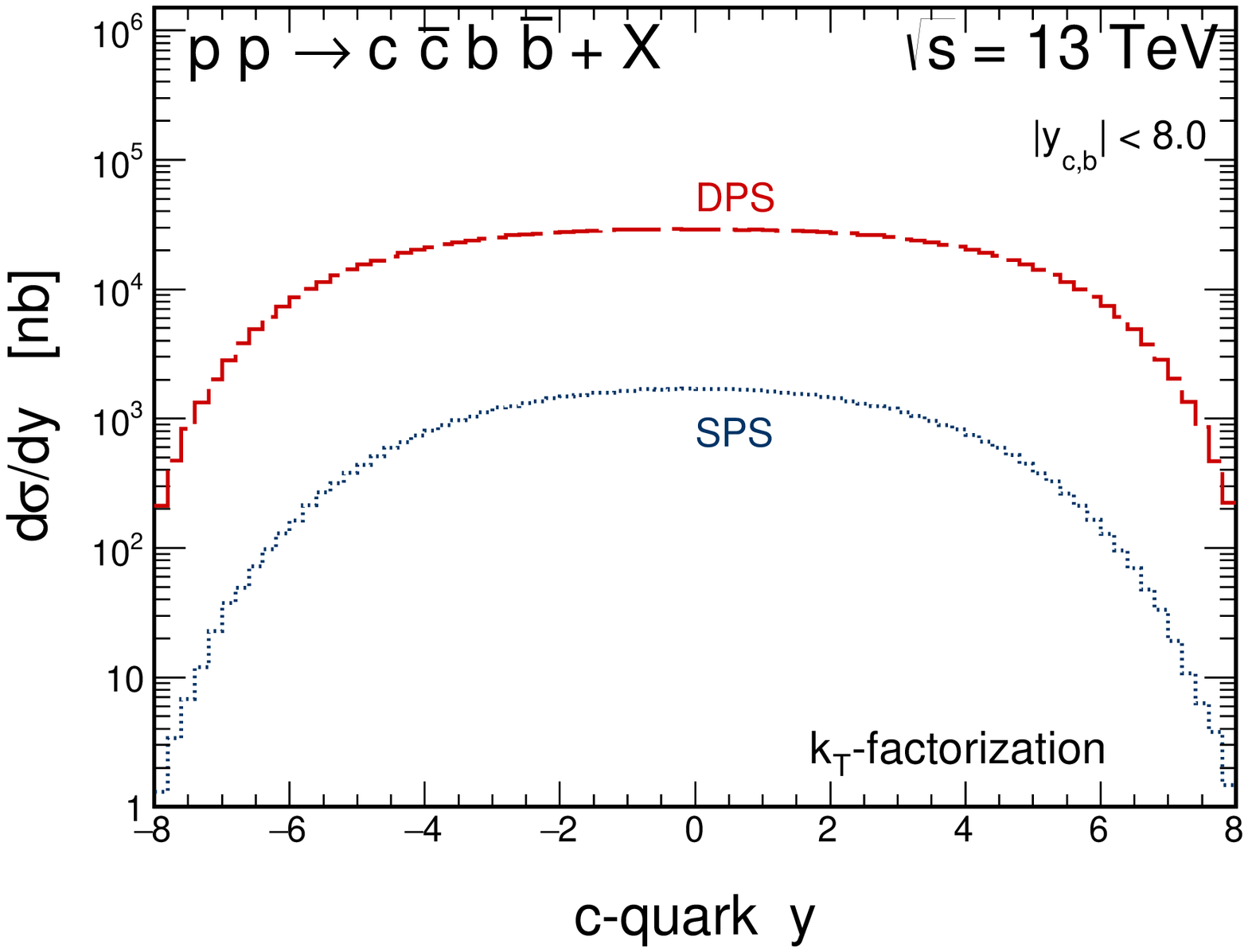}}
\end{minipage}
\hspace{0.5cm}
\begin{minipage}{0.47\textwidth}
 \centerline{\includegraphics[width=1.0\textwidth]{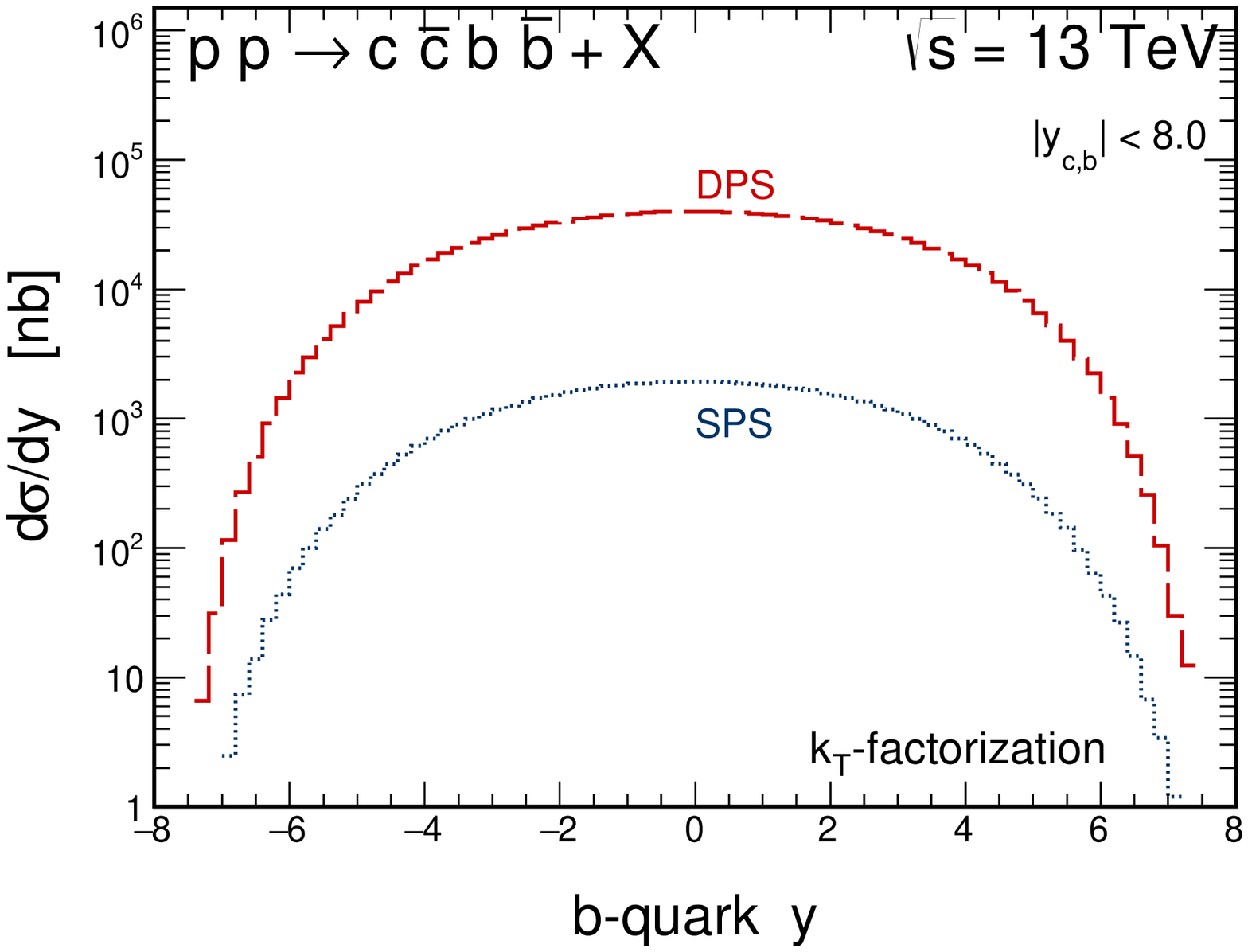}}
\end{minipage}
   \caption{
\small Transverse momentum (top) and rapidity (bottom) distributions of charm (left) and bottom (right) quark for the case of inclusive production of $c\bar c b\bar b$ final state. Contributions of the SPS (dotted) and the DPS (dashed) mechanisms are shown separately. The results are obtained within the $k_{T}$-factorization approach with the KMR uPDFs for $\sqrt{s} = 13$ TeV.
 }
 \label{fig:cb-pT-y}
\end{figure}

Now we go to the case of simultaneous production of charm and bottom particles. We start with the parton-level predictions for inclusive production of $c\bar c b\bar b$ final state at $\sqrt{s}=13$ TeV. In Fig.~\ref{fig:cb-pT-y} we show transverse momentum (top panels) and rapidity (bottom panels) distributions of charm (left panels) and bottom (right panels) quarks. The results are obtained for the full phase-space. The SPS (dotted histograms) and DPS (dashed histograms) contributions are shown separately.  
We observe that the DPS component significantly dominates over the SPS one in the whole rapidity range. It is also true for the transverse momentum distribution of bottom quark. In the case of charm quark the situation is slightly different. At small transverse momenta the DPS mechanism also gives dominant contribution, but both components become comparable when going to larger $p_{T}$'s.

\begin{figure}[!h]
\begin{minipage}{0.47\textwidth}
 \centerline{\includegraphics[width=1.0\textwidth]{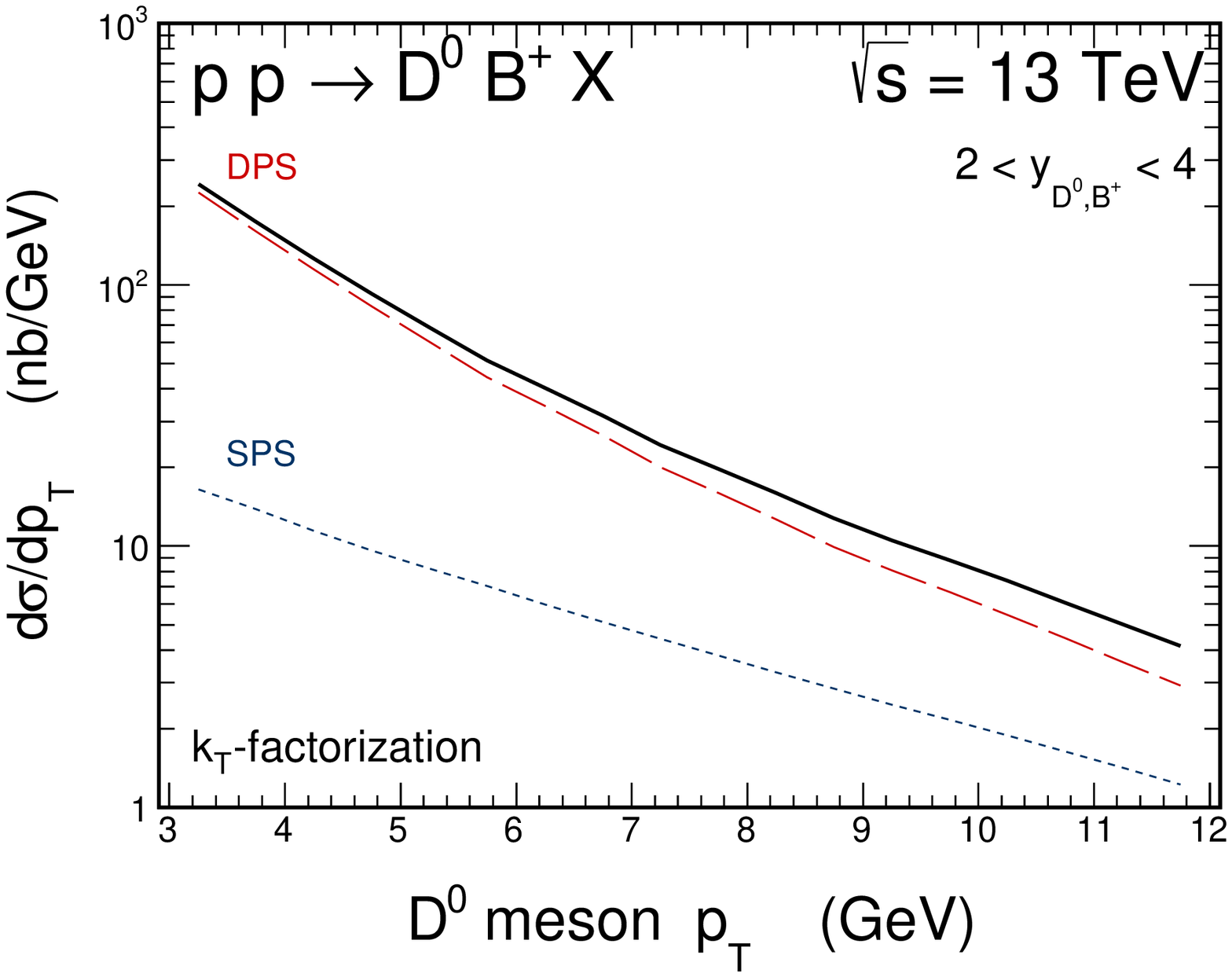}}
\end{minipage}
\hspace{0.5cm}
\begin{minipage}{0.47\textwidth}
 \centerline{\includegraphics[width=1.0\textwidth]{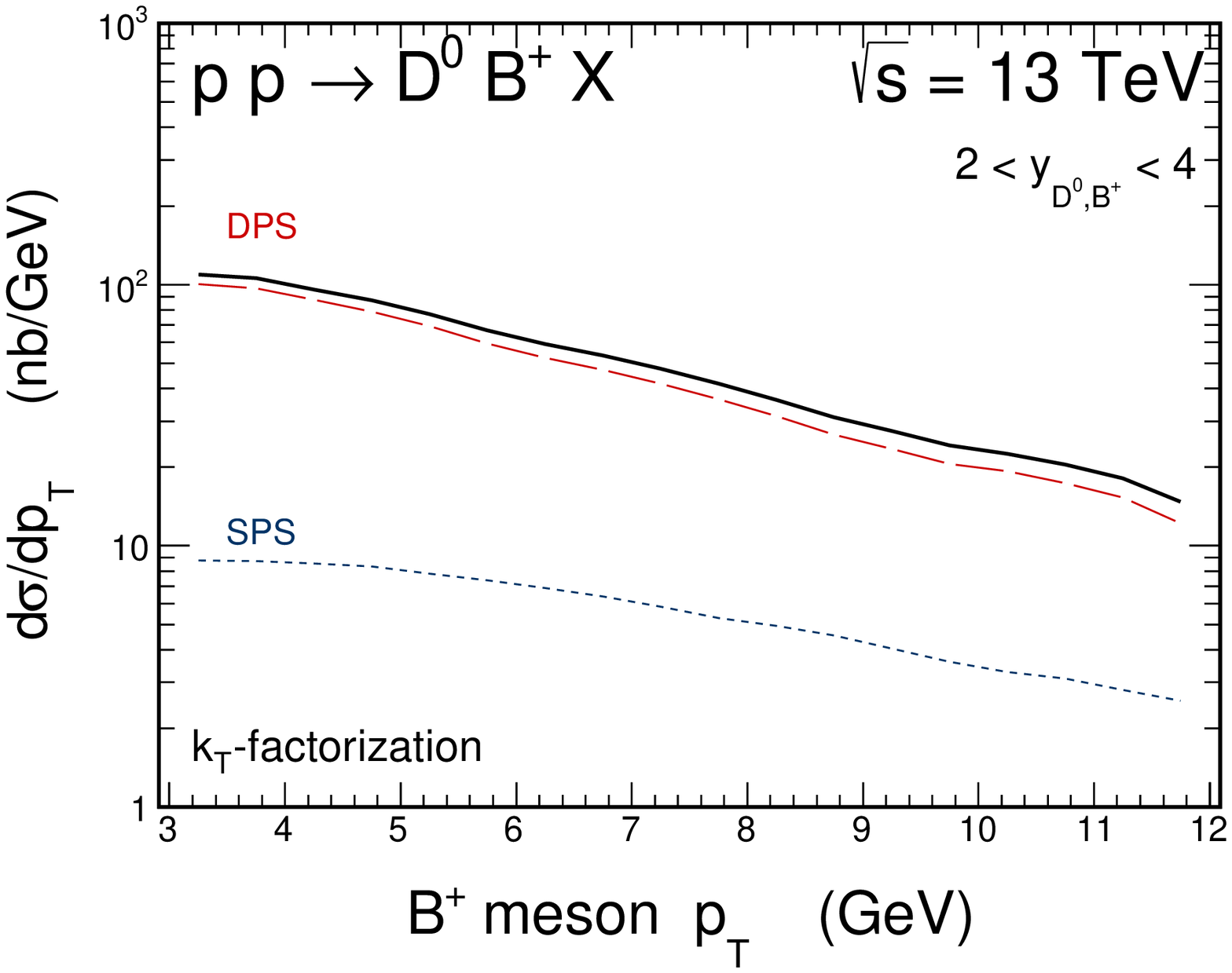}}
\end{minipage}
   \caption{
\small Transverse momentum distribution of $D^{0}$ (left) and $B^{+}$ (right) meson at $\sqrt{s} = 13$ TeV for the case of inclusive $D^{0}B^{+}$-pair production in the LHCb fiducial volume. The SPS (dotted) and the DPS (dashed) components are shown separately. The solid lines correspond to the sum of the two mechanisms under consideration. The results are obtained within the $k_{T}$-factorization approach with the KMR uPDFs. 
 }
 \label{fig:D0Bp-pT}
\end{figure}

The optimistic situation for searching for DPS effects in this channel presented above does not change when hadronization effects and
kinematical cuts relevant for the LHCb experiment are taken into account. We consider inclusive production of $D^{0}B^{+}$-pair since this mode has the most advantageous $cb \to DB$ fragmentation probability and leads to the biggest cross sections. In Fig.~\ref{fig:D0Bp-pT} we show the transverse momentum distribution of $D^{0}$ (left panel) and $B^{+}$ (right panel) meson at $\sqrt{s} = 13$ TeV for the case of simultaneous $D^{0}B^{+}$-pair production in the LHCb fiducial volume defined as $2 < y < 4$ and $3 < p_{T} < 12$ GeV for both mesons. Again, the SPS (dotted lines) and the DPS (dashed lines) components are shown separately, together with their sum (solid lines). Here, the conclusions are the same as for the parton-level results. We observe an evident enhancement of the cross section, at the level of order of magnitude, because of the presence of the DPS mechanism in the whole considered kinematical domain. We predict that the $D^{0}B^{+}$ data sample, that could be collected with the LHCb detector, should be DPS dominated in the pretty much the same way as in the case of double charm production (see \textit{e.g.} Ref.~\cite{Maciula:2013kd}).

\begin{figure}[!h]
\begin{minipage}{0.47\textwidth}
 \centerline{\includegraphics[width=1.0\textwidth]{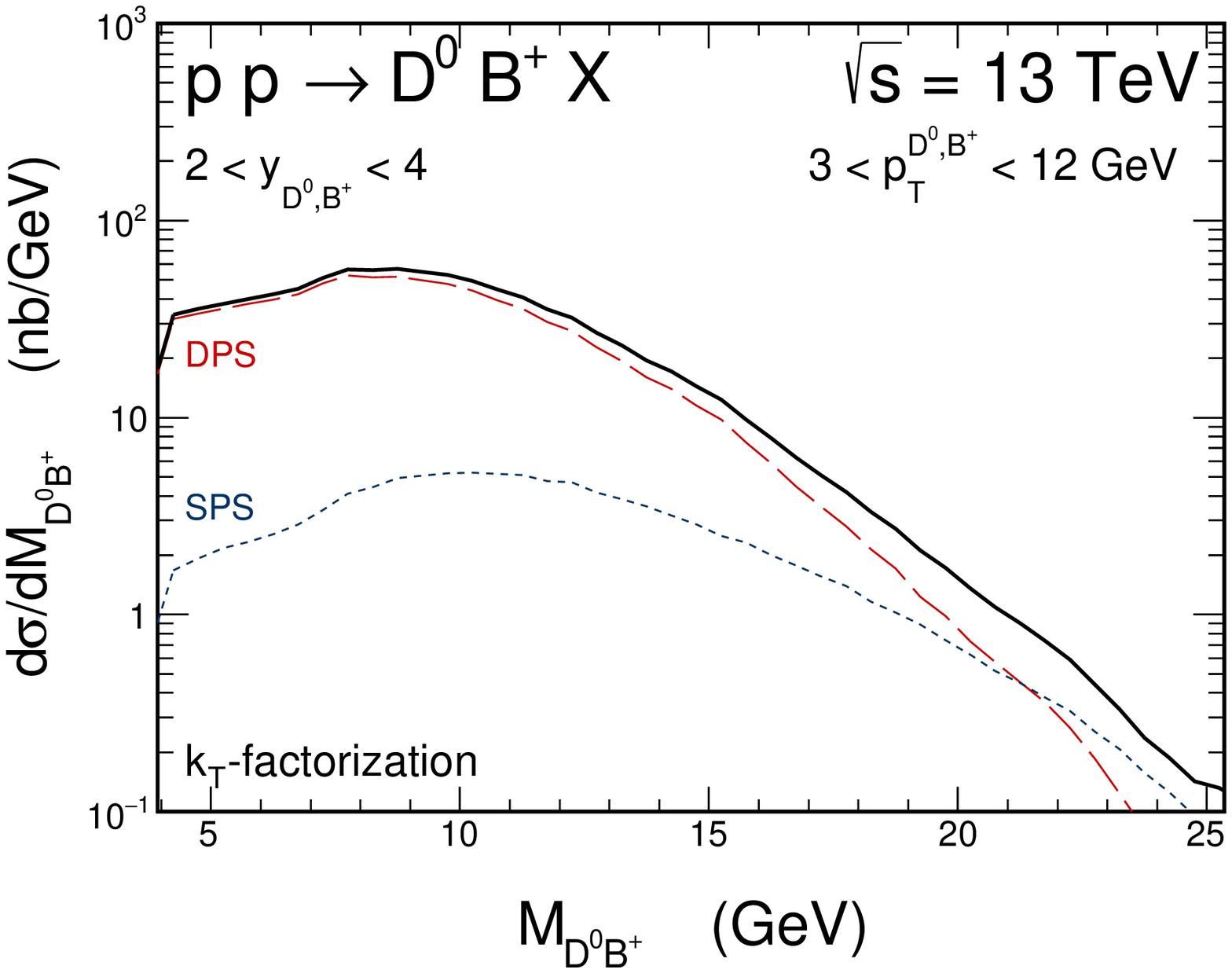}}
\end{minipage}
\hspace{0.5cm}
\begin{minipage}{0.47\textwidth}
 \centerline{\includegraphics[width=1.0\textwidth]{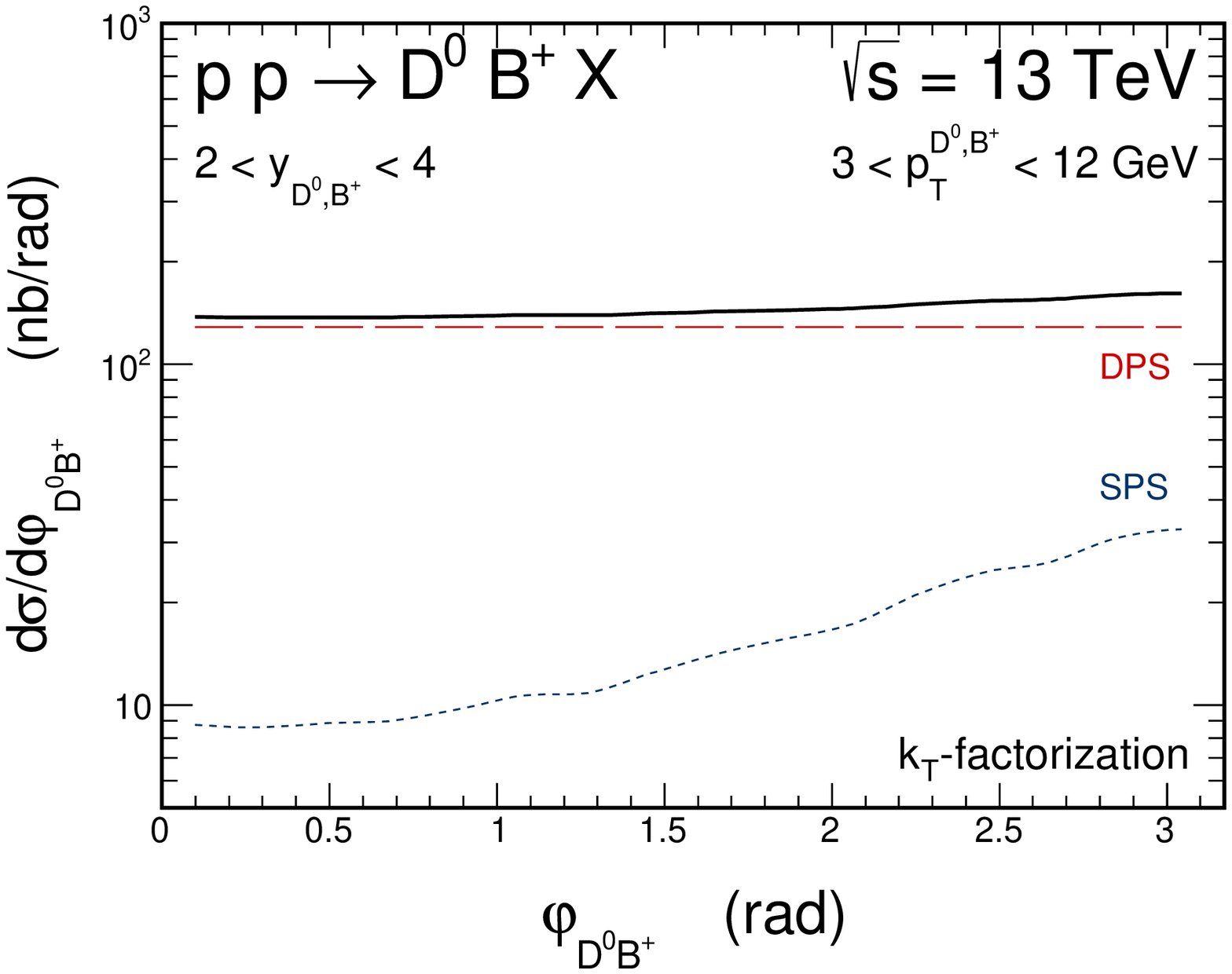}}
\end{minipage}
   \caption{
\small The same as in Fig.~\ref{fig:D0Bp-pT} but for the $D^{0}B^{+}$-pair invariant mass (left) and azimuthal angle $\varphi_{D^{0}B^{+}}$ (right) distributions.
 }
 \label{fig:D0Bp-corr}
\end{figure}

In Fig.~\ref{fig:D0Bp-corr} we present correlations observables that
could be helpful in experimental identification of the predicted DPS effects.
The characteristics of the di-meson invariant mass $M_{D^{0}B^{+}}$ 
(left panel) as well as of the azimuthal angle $\varphi_{D^{0}B^{+}}$ 
(right panel) differential distributions
is clearly determined by the large contribution of the DPS mechanism. 
We predict a significant enhancement of the cross section at small 
invariant masses $M_{D^{0}B^{+}} \lesssim 15$ GeV and a strong effect 
of azimuthal angle decorrelation, are related to the DPS mechanism. 

\begin{figure}[!h]
\begin{minipage}{0.47\textwidth}
 \centerline{\includegraphics[width=1.0\textwidth]{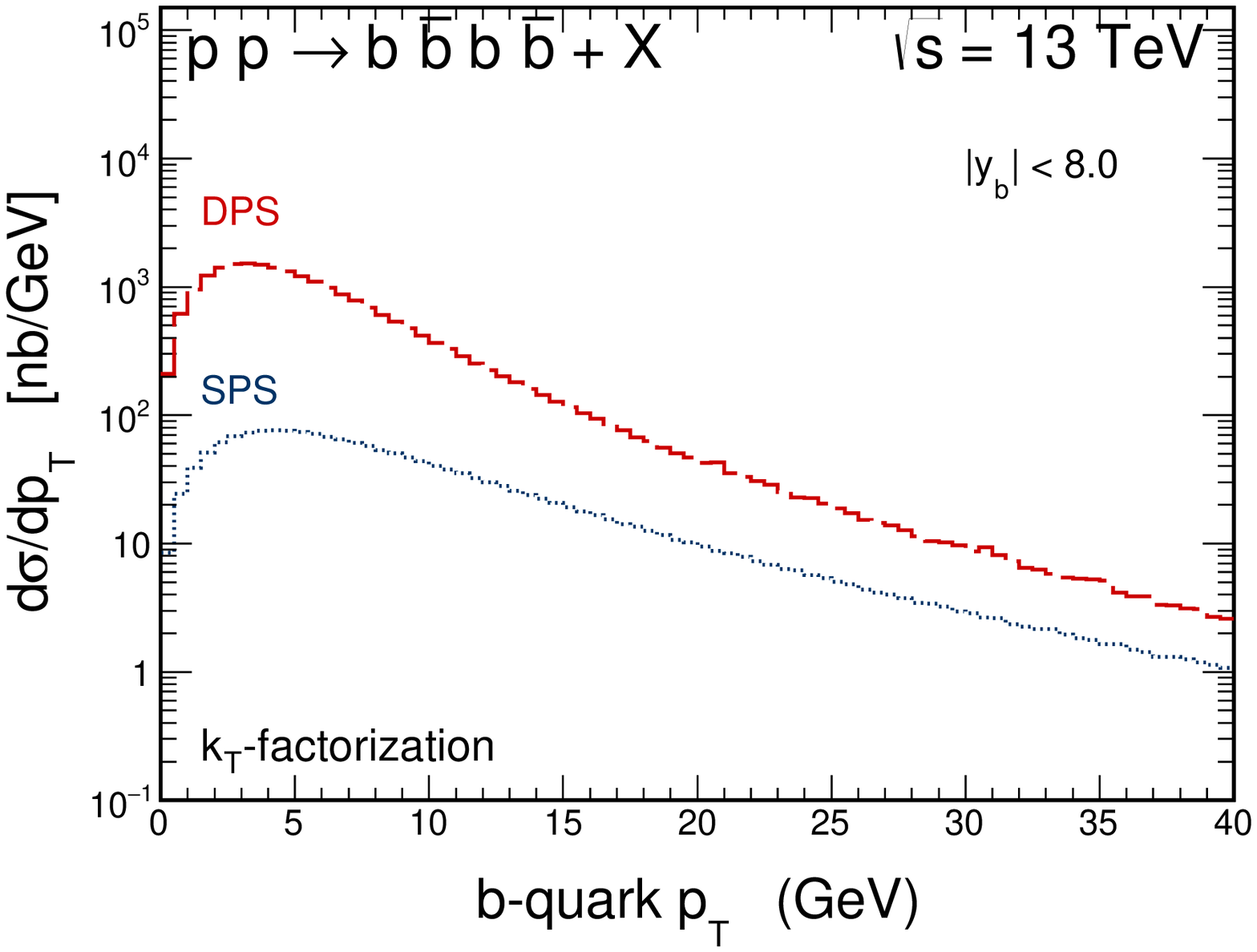}}
\end{minipage}
\hspace{0.5cm}
\begin{minipage}{0.47\textwidth}
 \centerline{\includegraphics[width=1.0\textwidth]{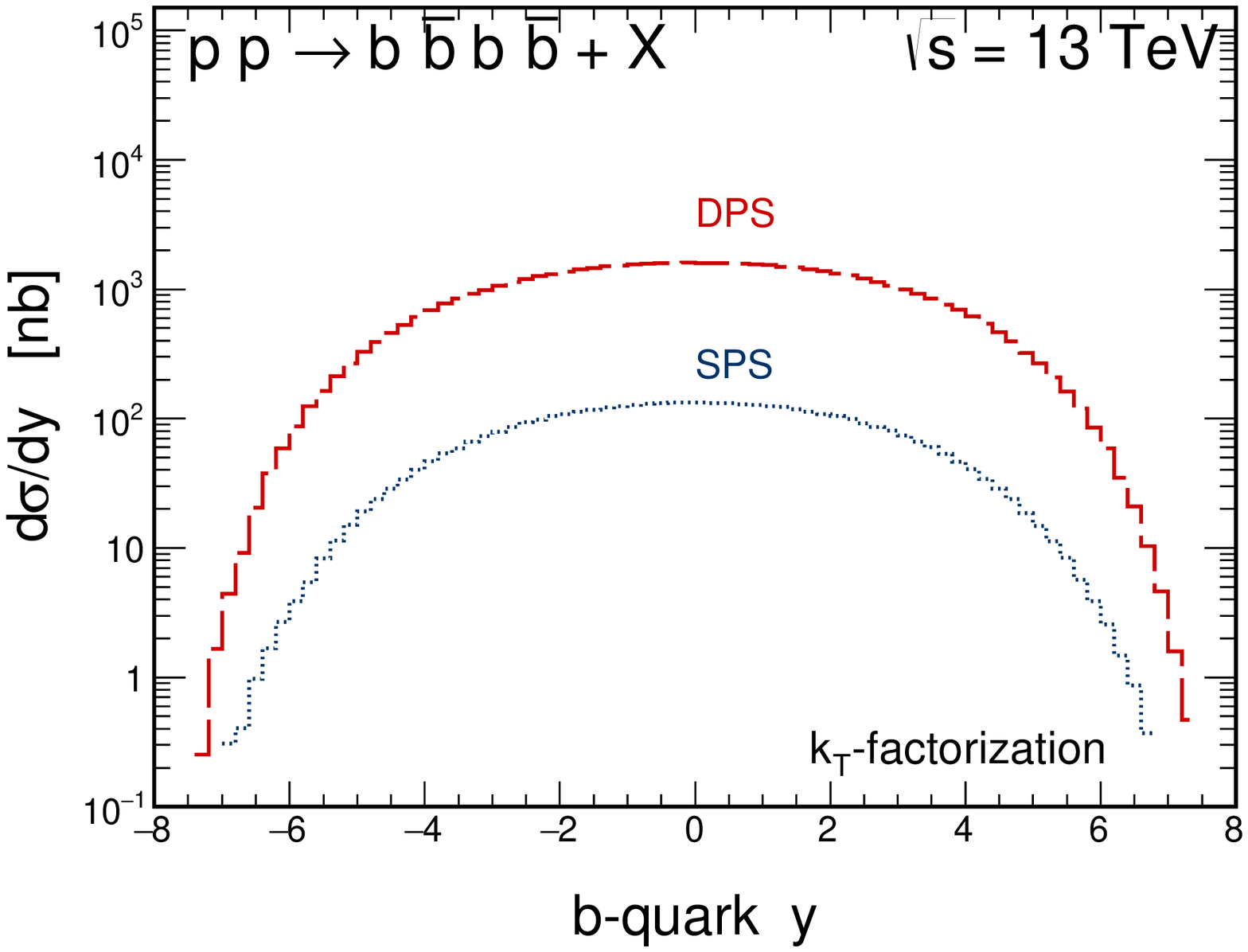}}
\end{minipage}
   \caption{
\small Transverse momentum (left) and rapidity (right) distributions of bottom quark for the case of inclusive production of $b\bar b b\bar b$ final state. Contributions of the SPS (dotted) and the DPS (dashed) mechanisms are shown separately. The results are obtained within the $k_{T}$-factorization approach with the KMR uPDFs for $\sqrt{s} = 13$ TeV.
 }
 \label{fig:bb-pT-y}
\end{figure}

Similar conclusions about a possibility of experimental observation 
and exploration of the DPS effects can be also drawn for the case 
of double bottom production.
As it is shown in Fig.~\ref{fig:bb-pT-y}, the relation between the SPS 
and the DPS components for the $b$-quark transverse momentum (left
panel) and rapidity (right panel) distribution in the case of 
$b\bar b b \bar b$ production is very similar to the relation 
predicted for the $c\bar c b \bar b$ final state (see right panels 
of Fig.~\ref{fig:cb-pT-y}). The main observed differences are the 
absolute normalization of the cross section, which is about order 
of magnitude smaller than in the case of $c \bar c b\bar b$, and 
a bit smaller relative contribution of DPS.

\begin{figure}[!h]
\begin{minipage}{0.47\textwidth}
 \centerline{\includegraphics[width=1.0\textwidth]{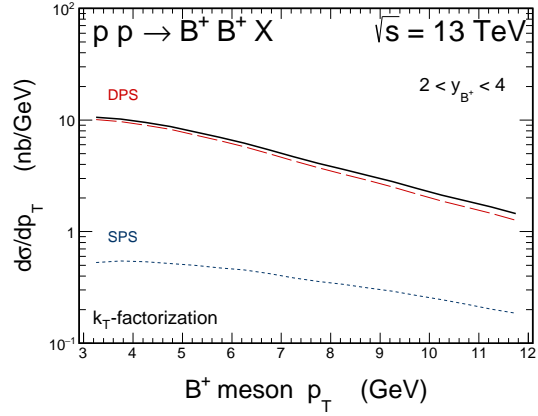}}
\end{minipage}
   \caption{
\small Transverse momentum distribution of $B^{+}$ meson at $\sqrt{s} = 13$ TeV for the case of inclusive $B^{+}B^{+}$-pair production in the LHCb fiducial volume. The SPS (dotted) and the DPS (dashed) components are shown separately. The solid lines correspond to the sum of the two mechanisms under consideration. The results are obtained within the $k_{T}$-factorization approach with the KMR uPDFs. 
 }
 \label{fig:BpBp-pT}
\end{figure}

\begin{figure}[!h]
\begin{minipage}{0.47\textwidth}
 \centerline{\includegraphics[width=1.0\textwidth]{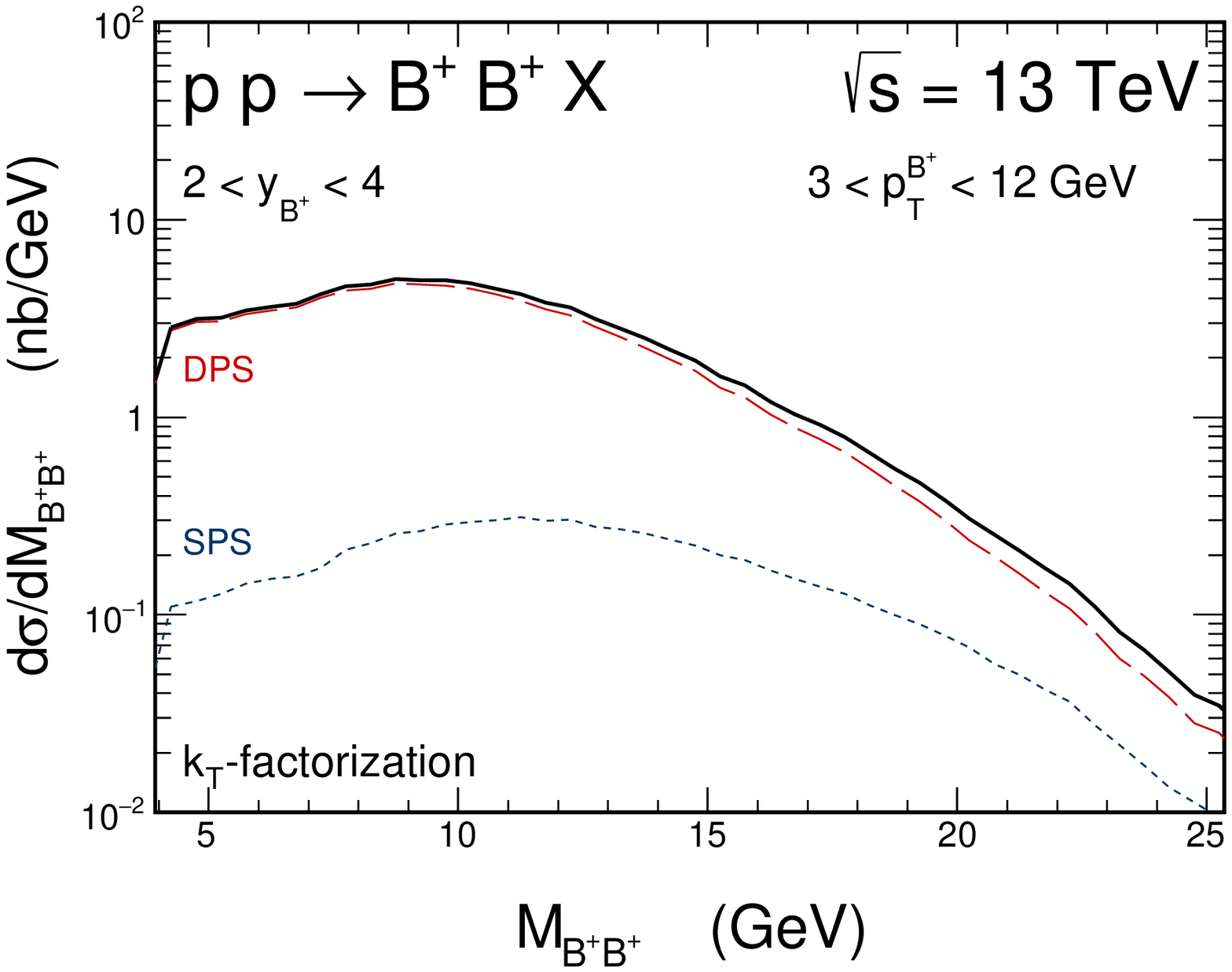}}
\end{minipage}
\hspace{0.5cm}
\begin{minipage}{0.47\textwidth}
 \centerline{\includegraphics[width=1.0\textwidth]{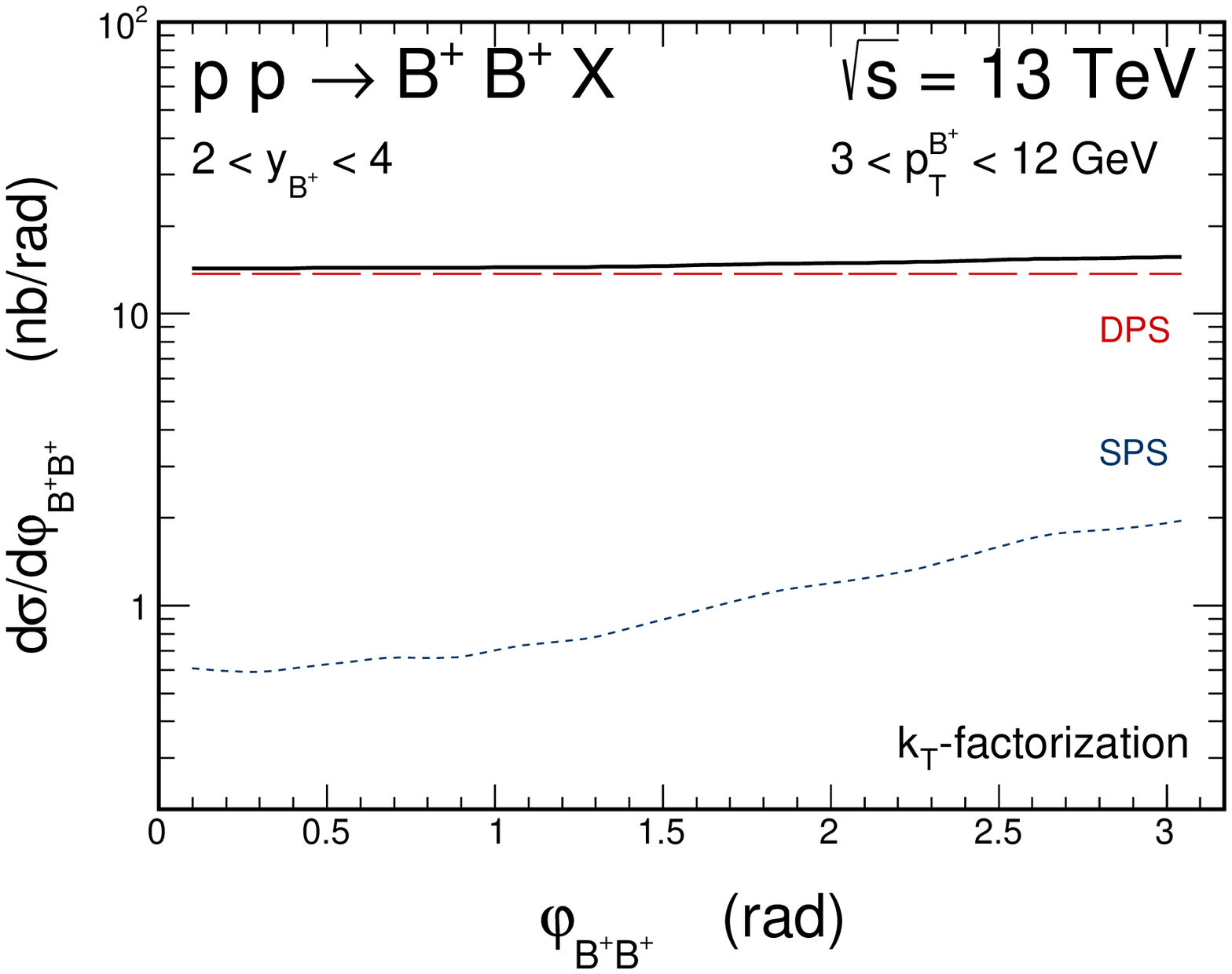}}
\end{minipage}
   \caption{
\small The same as in Fig.~\ref{fig:BpBp-pT} but for the $B^{+}B^{+}$-pair invariant mass (left) and azimuthal angle $\varphi_{B^{+}B^{+}}$ (right) distributions.
 }
 \label{fig:BpBp-Minv-Phid}
\end{figure}

The predictions for $B^+B^+$ meson-meson pair production for the LHCb
experiment only confirm the above statement. The effects related to 
the DPS mechanism on the $B^{+}$-meson transverse momentum 
(see Fig.~\ref{fig:BpBp-pT}),  on di-meson invariant mass 
$M_{B^{+}B^{+}}$  and on relative azimuthal angle $\varphi_{B^{+}B^{+}}$
(see left and right panels of Fig.~\ref{fig:BpBp-Minv-Phid})
distributions are pretty much the same as in the case of simultaneous
production of charm and bottom.

To summarize the situation for the LHCb experiment, in
Table~\ref{tab:total-cross-sections}, we collect the integrated cross
sections for $D^{0}B^{+}$ and $B^{+}B^{+}$ meson-meson pair production
in nanobarns within the relevant acceptance: $2 < y_{D^{0},B^{+}} < 4$
and $3 < p_{T}^{D^{0}, B^{+}} < 12$ GeV. We predict quite large cross
sections, in particular, at $\sqrt{s}=7$ TeV the calculated cross
section for $D^{0}B^{+}$ pair production is only 5 times smaller than
the cross section already measured by the LHCb for $D^{0}D^{0}$ final
state \cite{Aaij:2012dz}. The cross sections for $B^{+}B^{+}$ are order
of magnitude smaller than in the mixed charm-bottom mode, however, still
seems measurable.  
In both cases, the DPS component is the dominant one. The relative DPS
contribution for both energies and for both experimental modes is at the
very high level of 90\%.
This makes the possible measurements a very interesting from the point
of view of the multi-parton interaction community.  

\begin{table}[tb]%
\caption{The integrated cross sections for $D^{0}B^{+}$ and $B^{+}B^{+}$ meson-meson pair production (in nb) within the LHCb acceptance: $2 < y_{D^{0},B^{+}} < 4$ and $3 < p_{T}^{D^{0}, B^{+}} < 12$ GeV, calculated in the $k_{T}$-factorization approach. The numbers include the charge conjugate states.}
\label{tab:total-cross-sections}
\centering %
\newcolumntype{Z}{>{\centering\arraybackslash}X}
\begin{tabularx}{1.0\linewidth}{Z Z Z Z}
\toprule[0.1em] %

Final state & Mechanism  & $\sqrt{s} = 7$ TeV  & $\sqrt{s} = 13$ TeV       \\ [-0.2ex]

\bottomrule[0.1em]

\multirow{2}{3cm}{$D^{0}B^{+} + \bar{D^{0}}B^{-}$} & DPS  & 115.50  & 418.79 \\  [-0.2ex]
 & SPS  & 21.13  & 51.46 \\  [-0.2ex]
\hline
\multirow{2}{3cm}{$B^{+}B^{+} + B^{-}B^{-}$} & DPS  & 11.04  & 43.40 \\  [-0.2ex]
 & SPS  & 1.31  & 3.39 \\  [-0.2ex]
\bottomrule[0.1em]

\end{tabularx}
\end{table}

Now we wish to present results of similar studies as presented above but
for the CMS experiment. Here, the situation may be quite different than
in the case of the LHCb experiment because of the quite different
kinematical domains defined by the detector acceptance in both
experiments. The CMS experiment could collect the data for double bottom
production in the region of $|y_{B^{\pm}}| < 2.2$ and 
$10 < p_{T}^{B^{\pm}} < 100$ GeV. Here, crucial is the lower cut on 
meson transverse momenta which is quite large (much larger than in the
case of the LHCb). This may lead to damping of the relative DPS
contribution to the cross section under consideration.  

\begin{figure}[!h]
\begin{minipage}{0.47\textwidth}
 \centerline{\includegraphics[width=1.0\textwidth]{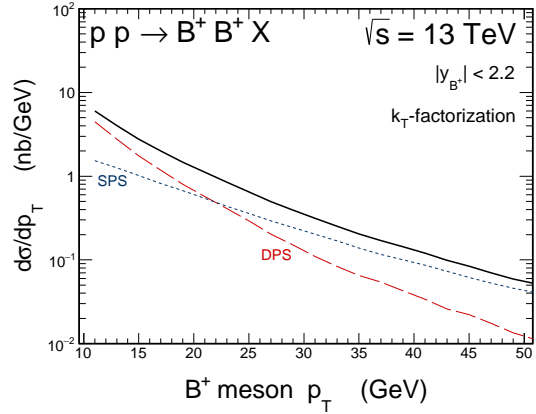}}
\end{minipage}
   \caption{
\small Transverse momentum distribution of $B^{+}$ meson at $\sqrt{s} = 13$ TeV for the case of inclusive $B^{+}B^{+}$-pair production for the CMS detector acceptance. The SPS (dotted) and the DPS (dashed) components are shown separately. The solid lines correspond to the sum of the two mechanisms under consideration. The results are obtained within the $k_{T}$-factorization approach with the KMR uPDFs. 
 }
 \label{fig:BpBp-pT-CMS}
\end{figure}

In Fig.~\ref{fig:BpBp-pT-CMS} we show the differential cross section as
a function of transverse momentum of $B^{+}$ meson for the CMS
experiment at $\sqrt{s} = 13$ TeV.
Here, the DPS mechanism dominates over the SPS one in the region of 
small transverse momenta $p_{T}^{B^{+}} \lesssim 20$ GeV, however, the
effect is not so strong as in the case of the LHCb experiment.

\begin{figure}[!h]
\begin{minipage}{0.47\textwidth}
 \centerline{\includegraphics[width=1.0\textwidth]{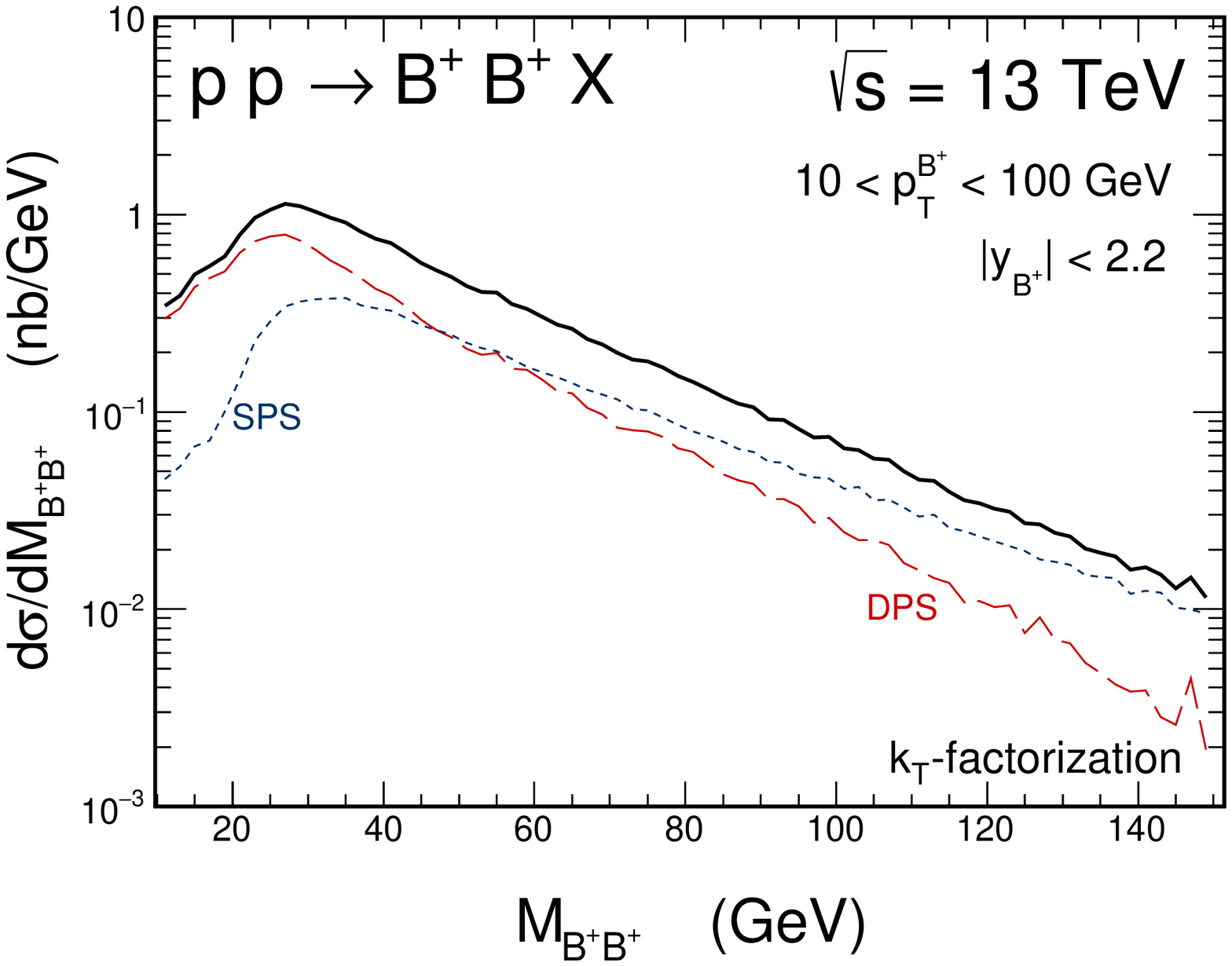}}
\end{minipage}
\hspace{0.5cm}
\begin{minipage}{0.47\textwidth}
 \centerline{\includegraphics[width=1.0\textwidth]{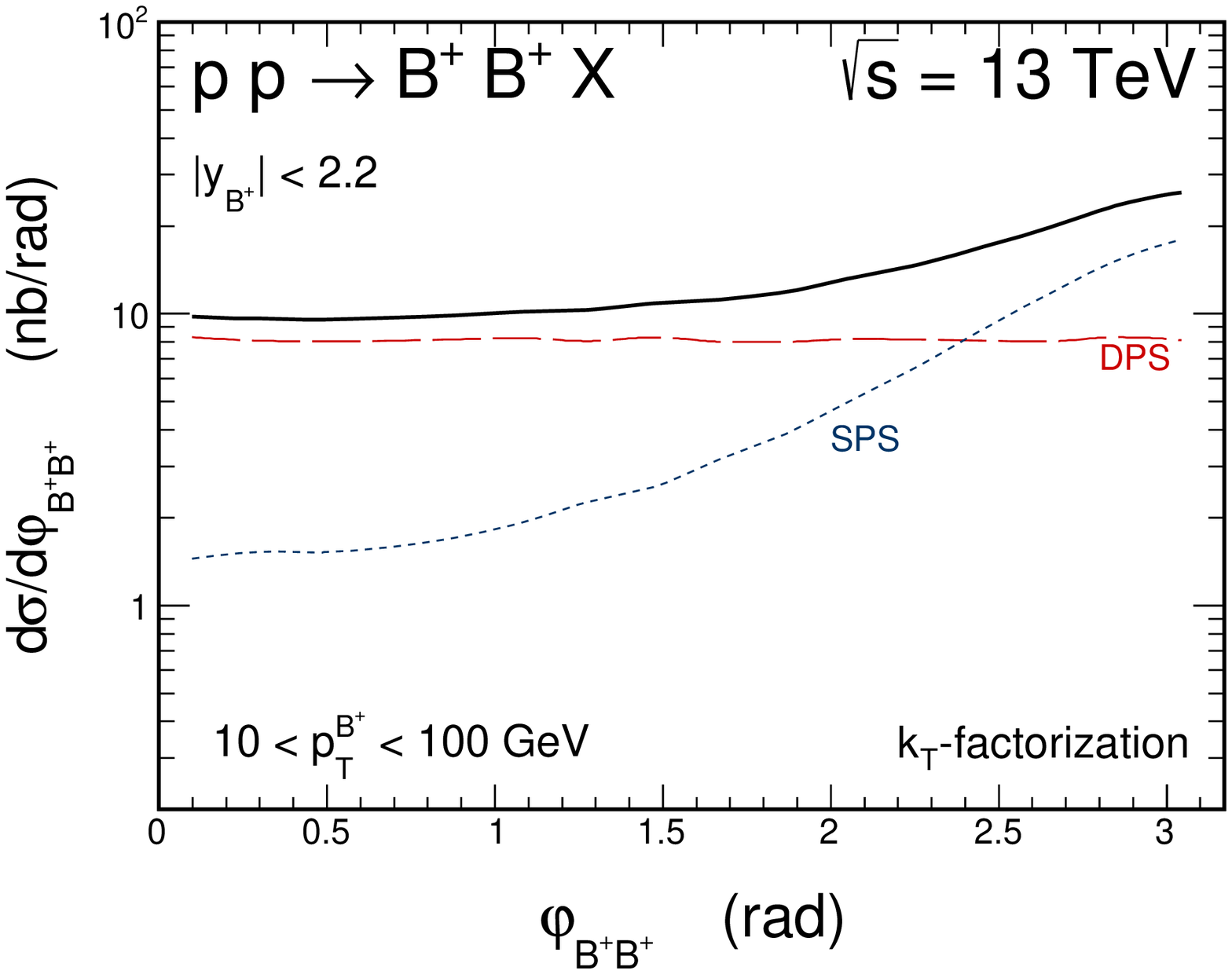}}
\end{minipage}
   \caption{
\small The same as in Fig.~\ref{fig:BpBp-pT-CMS} but for the $B^{+}B^{+}$-pair invariant mass (left) and azimuthal angle $\varphi_{B^{+}B^{+}}$ (right) distributions.
 }
 \label{fig:BpBp-Minv-Phid-CMS}
\end{figure}

In Fig.~\ref{fig:BpBp-Minv-Phid-CMS} we show the relevant distributions
in di-meson invariant mass $M_{B^{+}B^{+}}$ (left panel) and azimuthal 
angle $\varphi_{B^{+}B^{+}}$ (right panel). We observe a small effect of
the enhancement of the cross section especially at small invariant 
masses $M_{B^{+}B^{+}} \lesssim 50$ GeV, related to the DPS mechanism. 
The azimuthal angle $\varphi_{B^{+}B^{+}}$ distribution may be the most 
helpful for experimental identification of the DPS component within 
the CMS detector, since we predict also in this case a significant 
decorrelation of the distribution. 

Finally, in Table~\ref{tab:total-cross-sections_CMS} we show predictions for the integrated cross section. The calculated cross sections for $B^{+}B^{+}$ production
are very similar to those obtained for the LHCb detector, however, the relative DPS contribution for the CMS experiment is predicted at the level of 50\% and 60\% at $\sqrt{s}=7$ and $13$ TeV, respectively, i.e. smaller than in the case of LHCb.  

\begin{table}[tb]%
\caption{The integrated cross sections for $B^{+}B^{+}$ meson-meson pair production (in nb) within the CMS acceptance: $|y_{B^{\pm}}| < 2.2$ and $10 < p_{T}^{B^{\pm}} < 100$ GeV, calculated in the $k_{T}$-factorization approach. The numbers include the charge conjugate states.}
\label{tab:total-cross-sections_CMS}
\centering %
\newcolumntype{Z}{>{\centering\arraybackslash}X}
\begin{tabularx}{1.0\linewidth}{Z Z Z Z}
\toprule[0.1em] %

Final state & Mechanism  & $\sqrt{s} = 7$ TeV  & $\sqrt{s} = 13$ TeV       \\ [-0.2ex]

\bottomrule[0.1em]

\multirow{2}{3cm}{$B^{+}B^{+} + B^{-}B^{-}$} & DPS  & 6.84 & 26.27 \\  [-0.2ex]
 & SPS  & 7.24  & 17.05 \\  [-0.2ex]
\bottomrule[0.1em]

\end{tabularx}
\end{table}

\section{Conclusions}

In our previous studies we discussed in detail production of 
$c \bar c c \bar c$ and $c \bar c + \mathrm{2 jets}$ final states
in order to test and explore double-parton scattering effects.
In general the processes with charm production and/or jets with small
transverse momenta have large contribution of double-parton scatterings.
Here we have tried to complete the first stage of exploration of DPS effects
in the heavy flavour sector.

In the present paper we have extended our previous studies to
simultaneous production of $c \bar c$ and $b \bar b$
and two pairs of $b \bar b$. It was our aim to understand the interplay of
single- and double-scattering processes.
The calculation have been done within the standard so far factorized
ansatz with two independent partonic scatterings.
The so-called $\sigma_{\mathrm{eff}}$ parameter have been fixed at 
the same values as used in our previous studies for double charm production.
 
The cross section for each step has been calculated within the
$k_T$-factorization approach including transverse momenta of gluons entering
hard process. We have used the Kimber-Martin-Ryskin gluon distributions
that turned out so succesful for production of charm.
The hadronization of $c$ quarks to $D$ and $b$ quarks to
$B$ mesons have been done with the help of phenomenological
fragmentation functions. 
The Peterson fragmentation functions have been used.
We have obtained good description of the LHCb data 
for $B^0 + \bar B^0$ production with our standard choice of
factorization/renormalization scales. 

Having shown that inclusive $B$
meson transverse momentum distributions are rather well understood 
\footnote{The same was shown previously for $D$ mesons.} 
we have used our technique to calculate double parton
scattering processes. The calculation of double parton scattering have
been supplemented by calculation of single parton scattering (2 $\to$ 4)
processes using codes for automatized calculations of the off-shell
matrix elements, i.e. including transverse momenta of initial gluons.

First we have explored several different differential distributions
for $c \bar c b \bar b$ and $b \bar b b \bar b$ production for the whole
phase space.
We have observed clear dominance of the DPS over SPS for small transverse
momenta of $c$ or $\bar c$ and in the broad range of transverse momenta
of $b$ or $\bar b$.

Next we have considered distributions for simultaneous production of
charmed and bottom mesons. The DPS mechanism have been shown to dominate
for small invariant masses of the $D B$ systems. We have predicted only
a small decorrelation in relative azimuthal angle, typical for DPS
dominance.

The situation for $b \bar b b \bar b$ and two $B^+ B^+$ meson production
is rather similar as for the mixed heavy flavour production, but here 
the dominance of the DPS over SPS is limited to smaller corners of 
the phase space.
A good description of future data will therefore require to include 
both DPS and SPS mechanisms simultaneously.
All the considered reactions should be easily measured as the
corresponding cross sections are rahter large.

A comment on possible in principle measurements is in order.
Usually experimental subgroups specialize exclusively either in 
the production of $D$ mesons or $B$ mesons, simultaneous production 
of $D$ and $B$ mesons will require some coordination of the action of 
such different subgroups.
In our opinion it would be a valueble effort.
An experimental extraction of the $\sigma_{\mathrm{eff}}$ parameter for different
reactions and a comparison for different processes studied here
and in our previous papers would be a simple but necessary step
to better understand double scattering in a more precise way.
Also a compilation of the $\sigma_{\mathrm{eff}}$ would be important
phenomenological knowledge.
The factorized ansatz is an approximation and a possible deviations
from it were discussed in the literature.
Once such studies as discussed here are completed one can try 
to explore deviations from the simple approach. No clear deviations were
found so far.
The only exception is production of quarkonia pairs were very small
values of $\sigma_{\mathrm{eff}}$ were extracted from experimental data.
The situation in quarkonia pair production is however more complex.
As disussed recently in Ref.~\cite{CSS2017} there are several single-parton mechanisms
with DPS characteristics. Such processes were not considered so far
in theoretical calculations so the extraction of $\sigma_{\mathrm{eff}}$
for these reactions is not reliable. 
Therefore in DPS studies one should concentrate first rather on
processes with heavy quark/meson production.

\vspace{1cm}

{\bf Acknowledgments}

This study was partially
supported by the Polish National Science Center grant
DEC-2014/15/B/ST2/02528 and by the Center for Innovation and
Transfer of Natural Sciences and Engineering Knowledge in
Rzesz{\'o}w.


\end{document}